\documentstyle[preprint,aps,eqsecnum]{revtex}

\input psfig
\begin{document}
\draft
\preprint{FERMILAB-Pub-95/220-A}
\baselineskip.5cm
\parskip4pt
\date{\today}
\title{Condensation of bosons in kinetic regime}
\author{D. V.\ Semikoz$^{1}$ and I. I.\ Tkachev$^{1,2}$}
\address{
$^{1}$Institute for Nuclear Research of
Russian Academy of Sciences, Moscow 117312, Russia}
\address{
$^{2}$NASA/Fermilab Astrophysics Center
Fermi National Accelerator Laboratory, Batavia, IL 60510}

\maketitle

\begin{abstract}
\baselineskip0.5cm
We study the kinetic regime of the Bose-condensation of scalar particles with
weak $\lambda \phi^4$ self-interaction. The Boltzmann equation is solved
numerically. We consider two kinetic stages. At the first stage the condensate
is still absent but there is a nonzero  inflow of particles towards ${\bf p} =
{\bf 0}$ and the distribution function at ${\bf p} ={\bf 0}$ grows from finite
values to infinity in a finite time.
We observe a profound similarity between Bose-condensation and Kolmogorov
turbulence. At the second stage there are two components, the condensate and
particles, reaching their equilibrium values. We show that the evolution in
both stages proceeds in a self-similar way and find the time needed for
condensation. We do not consider a phase transition from the first stage to the
second. Condensation of self-interacting bosons is compared to the condensation
driven by interaction with a cold gas of fermions; the latter turns out to be
self-similar too. Exploiting the self-similarity we obtain a number of
analytical results in all cases.
\end{abstract}

\narrowtext

\section{Introduction}
\label{sec:In}
It is a fundamental result of quantum statistics that above a certain critical
density all added bosons must enter the ground state: a Bose-Einstein
condensate (BEC) forms. Gas of weakly interacting bosons allows a well-defined
theoretical microscopic treatment of this process. However, Bose condensation
has not yet been observed in a real physical system which could be described as
an ideal gas. Superfluid $^4$He, for example, is a system with very strong
interaction and is considered as a quantum liquid.

One can reach Bose condensation by gradually decreasing the temperature in a
sequence of equilibrium states.  An appropriate description of this regime is
given by the well-known theory of second order phase transitions \cite{ll}. On
the other hand, the conditions for the formation of a Bose condensate can
appear when the system is far from equilibrium. This will be generally true for
the condensation of an ideal gas because of the weakness of interaction. The
kinetics of Bose-condensation in this case is an important and interesting
problem.

Recently this subject has attracted particular interest in connection with the
exciting prospects for experimental observation of a Bose condensation in very
cold atomic samples, {\it e.g.} in a gas of spin-polarized atomic hydrogen
\cite{sw86,bprl} or in alkali-metal vapors \cite{m90}, as well as in the  gas
of excitons (bound states of electrons and holes in semiconductors)
\cite{exbc}. For instance, Doyle {\it et. al.} \cite{bprl} report to have
attained $n \approx 10^{14}$ cm$^{-3}$ and $T  \approx 100 ~\mu{\rm K}$, which
is only a factor of 3 above the temperature of Bose condensation at this
density.

Since the life-time of a sample in all experiments is finite and short, the
question of a time-scale for the formation of Bose-condensate is a central one.

Another interesting application of Bose kinetics is rather far from laboratory
experiments and is related to the problem of Bose star \cite{bs}  formation
\cite{it91} from the dark matter in the Universe. The life-time of the
``sample'' is not limited in this case, but in the only model \cite{kt93} for
Bose-star formation which involves a realistic dark matter particle candidate,
the axion \cite{mt90}, the self-interaction of particles is so weak that the
relaxation time is comparable to the age of the Universe. The possibility of
Bose-star formation becomes dependent upon the exact value of this time-scale.

In experiments with atomic hydrogen or with alkali-metal vapors the Bose-gas is
effectively isolated from its surroundings and the relaxation to thermal
equilibrium and condensation is expected to occur due to particles'
self-coupling only. The same is true for the case of axion miniclusters. On the
contrary, interaction with the thermal heat bath is important in the dynamics
of excitons. We consider both situations in the present paper. However, we only
consider the simplest $\lambda \phi^4$ self-interaction of structureless scalar
particles (this approximation is exact {\it e.g.} in the axion case). The heat
bath is modelled by a gas of heavy fermions.

The process of Bose condensation can be divided into three stages. The first
and the third ones are kinetic stages which occur before and after
actual nucleation of the condensate. These two stages can be treated with
the Boltzmann equation which describes restructuring of the distribution
function at the first stage and growth of the condensate at the third.
A numerical study of these two stages is the purpose of the present
paper (we have reported the main results in Ref. \cite{st95}).

In the framework of the kinetic equation there is no condensate at all
times if there was no condensate initially. Therefore we have  to add a small
seed condensate ``by hand'' in this framework when switching from the first
to the third stages. We switch between stages when the distribution function
becomes infinite at zero momentum. We do not consider the second
stage, {\it i.e.} how the condensate actually emerges, which is, in fact, a
phase transition. A description of the second stage has been elaborated in Ref.
\cite{s91}. Since the duration of this stage is short and the magnitude of the
condensate which appears is small \cite{s91}, our approach seems to be
reasonable.

The question of the time evolution of a weakly interacting Bose gas during the
first kinetic stage was addressed in a number of papers. In earlier treatments
an ideal Bose gas was coupled to a thermal bath with infinite heat capacity: to
a phonon bath in Ref. \cite{ih76} and to a fermi heat bath in Ref. \cite{ly77}.
A small energy exchange was assumed and after several other approximations a
Fokker-Planck type of equation was obtained. Levich and Yahkot \cite{ly77}
calculated analytically that the time for condensation (more precisely, the
time needed for the distribution function to reach infinity at zero momentum)
is infinite in this situation. By including boson-boson interactions they later
\cite{ly78} found a solution which describes the explosive emergence of a
condensate, but they concluded that this effect could have been an artifact of
their approximations.

Snoke and Wolfe in Ref. \cite{sw89} undertook a direct numerical integration of
the Boltzmann kinetic equation. Although this calculation demonstrated a
restructuring of the distribution of particles, the appearance of a Bose
condensate (more precisely, an infinite value of the distribution function at
zero momentum) was not detected. As compared with Ref. \cite{sw89}, we carry
out the numerical integration in a much wider dynamical range of relative
energies and densities. The major and important difference of our approach is
that we directly analyse the behavior of the distribution function, while Snoke
and Wolfe considered an integrated quantity $n(\varepsilon)$, the number of
particles with the energy less than a given value $\varepsilon$.

Analytical study of Bose condensation was performed recently by Kagan,
Svistunov and Shlyapnikov in the paper \cite{kss92} where three different
regimes of evolution were identified and considered. Our numerical results
confirm qualitative predictions of that work. In particular, it was argued that
in the kinetic region of the non-linear regime the distribution function has
the form $f(\varepsilon) \sim \varepsilon ^{-7/6}$ for small $\varepsilon$.  We
indeed observe the tendency to this law in our numerical integration, but,
while being close to it, the system never reaches this critical exponent.
Instead of $7/6 \approx 1.17$ we obtain 1.24. Moreover, this distribution is
destroyed when the condensate emerges, the fact that was not stressed in
Ref.\cite{kss92}. As concerns the condensation time, only order of magnitude
dimensional estimates were obtained in Refs. \cite{it91,kss92}.

We find that during the first kinetic stage the distribution function became
infinite at zero momentum in a finite time in the case of self-interacting
bosons. This stage can be divided into two parts. During the first part of the
first stage the distribution function is restructured so that a self-similar
solution is formed; the distribution function does not grow much during this
time interval which takes about half of the entire evolution time. During the
second part, the distribution function rapidly (as a pole power law $\propto
(t_c-t)^{-2.6}$) reaches infinity at ${\bf p}=0$ in a self-similar way.

Despite the fact that the Bose-condensation is inherently based on
Bose-statistics, the kinetics of this stage is a classical process and not a
quantum one. To be precise, it corresponds to the kinetics of classical waves
(but not classical particles). Actually, the equations we encounter here and
the resulting behavior of the system are well-known in the theory of turbulence
\cite{my,zmr85,zlf}. This is no wonder. Indeed, a Bose condensation occurs when
the occupation numbers are large, $f({\bf p}) \gg 1$. In that case we can
replace quantum creation operators by amplitudes of classical waves. The
corresponding kinetic equation can be obtained in two ways: either neglecting
$1$ in the product of all factors $[1+f({\bf p_i})]$ in the quantum kinetic
equation (by which we neglect the spontaneous scattering  with respect to the
induced one), or directly in the random phase approximation for the ensemble of
classical waves \cite{zmr85,zlf}. The second approach is customary in the
theory of turbulence, but the result is the same in both
approaches.\footnote{\footnotesize \baselineskip0.3cm This is particularly
important for the axion case, because axions in miniclusters \cite{kt93} are
from the very beginning in a state of gravitationally bounded classical waves
and one may wonder (or be suspicious of) how Bose-condensation could occur in
such a system.}
While it may be more appropriate in the second approach to refer to
$f(\varepsilon)$ as a density of waves in phase space, we will continue to call
it density of particles in phase space.

The power law behavior of the distribution function we observe, with an
exponent close to $7/6$, is a special case of the Kolmogorov turbulence
\cite{k41,za66,za76}. Indeed, by definition, turbulence is a stationary state
characterized by a non-zero flow (in phase space) of some conserved quantity.
In the case of Bose-condensation we have a flux of particles towards the
condensate, which tends to be constant across the energy space, and this
corresponds to $f(\varepsilon) \propto \varepsilon^{-7/6}$. An explosive
behavior of the distribution function at ${\bf p}=0$ which we observe was also
found in other instances of turbulence \cite{fs91,zlf}.

The actual nucleation of the condensate cannot be described in the framework of
the kinetic equation. But we have to conclude that after the distribution
function has become infinite at zero momentum, the condensate forms. Kinetic
equations which describe the third stage in the presence of the condensate were
derived in Refs. \cite{e84}. We re-derive those equations in a simpler way
which is valid when the difference between particles and quasiparticles is
unimportant. We consider this weak coupling regime only. We found that during
the third stage the evolution is also self-similar until the restructuring in
the phase-space reaches the tail of the distribution.

We compare this condensation process, which is due to self-interaction of
particles, to the condensation driven by the interaction with a cold bath, for
which we also solved the Boltzmann equation numerically. We find
that the distribution function grows with time only as $\propto t^3$ in this
case. The energy dependence of the distribution function is also different now.
Instead of a singular power law profile, $f(\varepsilon) \sim \varepsilon
^{-7/6}$, which corresponded to the constant flux of particles towards
the condensate, now all excess particles form a well defined  packet in the
momentum space. The total number of particles in this packet remains constant
during the evolution, while its width decreases, gradually approaching a
$\delta$-function \cite{ly77}. This type of behavior is qualitatively different
from the one we have in the case of self-interacting bosons. However, it is
also self-similar and one of the predictions of the self-similarity in this
case is that $f(\varepsilon=0,t) \propto t^3$.  In general, using the
properties of the self-similarity enables us to obtain a number of analytical
results, {\it e.g.} the relation of the critical exponent $\beta$ in
$f(\varepsilon=0,t) \propto (t_c-t)^{-\beta}$ to the critical exponent $\alpha$
in $f(\varepsilon) \propto \varepsilon^{-\alpha}$.

The fact that the system interacting with a cold bath needs infinite time (in
the kinetic framework) to reach infinity in the distribution function at ${\bf
p} = 0$ does not  necessarily mean that the condensation time is infinite for
this system. Indeed, the nucleation of the condensate is a phase transition
which does not have to occur at the moment when the distribution reaches
infinity, and can occur earlier.

Although our prime interest and motivation for this work were connected
with the physics of Bose stars, the problem of condensate formation in a
gravitational field has never been considered, and we do not
consider it here. Instead, we are solving kinetic equations in flat
space-time and only the range of parameters of the initial distribution
can reflect the virial equilibrium of a self-gravitating system.

When this work was completed B. Svistunov brought to our attention his paper
\cite{bs91}. He also analyzed two non-linear kinetic stages exploiting the
assumption of self-similarity.
His approach is close to ours and our results qualitatively confirm the results
and predictions of Ref. \cite{bs91}. However, he assumed a priory in his
analysis that the critical exponent is $\alpha=7/6$. He concluded that this
assumption had passed his numerical self-consistency check, but we believe that
this is due to the fact that he used energy grid with the minimum value
$\varepsilon_{\rm min} =10^{-3}$, while in our case it was $\varepsilon_{\rm
min} =10^{-9}$.

The plan of this work is as follows. In Sec.~\ref{sec:Ki} we reduce the
Boltzmann kinetic equation  to the form suitable for numerical integration,
both in the absence and presence of the condensate.
In Sec.~\ref{sec:NC} we present the results of numerical integration which
describe the evolution of one-particle distribution function in the case of
self-interacting bosons, while
Sec.~\ref{sec:zeroT} is devoted to the evolution of a Bose system
interacting with fermion bath of infinite heat capacity.
In Sec.~\ref{sec:ss} we derive analytical consequences of self-similar behavior
of the distribution function.
In the final section \ref{sec:Con} we discuss some possible applications  and
present conclusions.
Appendix \ref{sec:Simp} describes an analytical reduction of the collision
integral. In Appendix \ref{ssec:zero} the static (``turbulent'') power law
solutions to the kinetic equation are described.
Appendix \ref{zT} is a technical supplement to Sec.~\ref{sec:zeroT}.

\section{The kinetic equation}
\label{sec:Ki}
\begin{figure}
\psfig{file=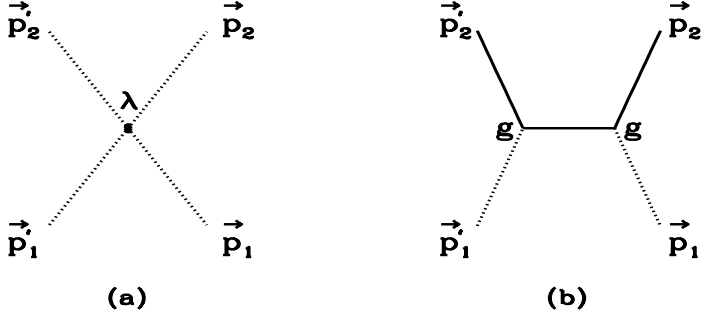,height=5cm,width=10cm}

\caption{The scattering processes we consider in
Eq. (2.1). (a) is the case of self-interacting bosons, and (b) is the case of
ideal Bose gas interacting with the cold fermion bath.}
\label{fig1}
\end{figure}

In a state of thermal equilibrium any system of bosons has to have the
Bose-Einstein distribution of particles over the energies. In order to reach
this
state particles need to interact either with the thermal bath of other
particles or with each other.
We shall consider both cases: the system of scalar bosons with
4-particle self-interaction and an ideal Bose gas coupled to cold gas of
fermions. The time development of a state which contains
large number of particles can be
adequately described by the kinetic equation which governs the
evolution of the one-particle distribution function, $f({{\bf p}})$.
The only process which contributes to the Boltzmann kinetic equation
is illustrated in Fig. \ref{fig1} for each case we consider, and the equation
takes the form:
\begin{equation}
\frac{df ({{\bf p}}_1)}{dt} = \frac{(2\pi)^4}{2p_{01}} \int
\delta^4(\sum_k p_{\mu k}) |M_{fi}|^2 F(f) \prod_{k=2}^4
\frac{d^3{{\bf p}}_k}{(2\pi)^3 2p_{0k}}\, ,
\label{kin}
\end{equation}
where $(p_{0k},{{\bf p}}_k)$ are components of the four-vector $p_{\mu k}$. For
the conceptual convenience we wrote this equation in the most general
Lorentz-invariant form, but in what follows we shall consider only the
non-relativistic limit of it.
For the quartic self-interaction, Fig. \ref{fig1}a, we have
\begin{mathletters}
\label{Fun}
\begin{equation}
F(f)=[1+f_1][1+f_2]f'_1f'_2-[1+f'_1][1+f'_2]f_1f_2,
\label{Fun1}
\end{equation}
where $f_k \equiv f({\bf p}_k)$, $f'_k \equiv f({\bf p}_k\, ')$ (we assume the
convention $1'=3$, $2'=4$), while
for the interaction with the fermion bath, Fig 1b :
\begin{equation}
F(f)=[1+f_1][1-\eta^{}_2]f'_1\eta'_2-[1+f'_1][1-\eta'_2]f^{}_1\eta^{}_2,
\label{Fun2}
\end{equation}
\end{mathletters}
where $\eta^{}_k \equiv \eta({\bf p}_k)$ is the one particle distribution
function of fermions which we shall not evolve here and consider it to be fixed
to the equilibrium.

The field-theoretical Lagrangian which we bear in mind for the case of
self-interacting bosons is
\begin{equation}
{\cal L} = \frac{1}{2}\partial_\mu \phi^2 + \frac{m^2}{2}\phi^2 +\lambda
\frac{\phi^4}{4!}~.
\label{lag}
\end{equation}
and the corresponding matrix element is given by
\begin{equation}
|M_{fi}|^2=  \lambda^2 ~.
\label{Mfi}
\end{equation}
The matrix element of the process in Fig. \ref{fig1}b  in the non-relativistic
limit is also momentum independent, $|M_{fi}|^2=g^4$.

The Bose-Einstein distribution function is
\begin{equation}
f({{\bf p}})=\frac{1}{\exp{[(\varepsilon-\mu)/T]}-1} + (2\pi)^3 n_c \delta
({{\bf p}}) ~,
\label{stat}
\end{equation}
where $\varepsilon$ is the particle energy, $\mu$ is the chemical potential,
$T$ is the temperature of the final state  and $n_c$ is the density of
particles in the condensate.
The fact that distribution function (\ref{stat}) is a solution to
the equation
$df({{\bf p}}_1)/dt \equiv I_{\text{coll}}({{\bf p}}_1)=0$ is a simple
consequence of the energy-momentum
conservation in each particle interaction vertex and does not depend upon the
specific form of the matrix element $M_{fi}$.

One expects that an arbitrary initial distribution function $f({{\bf p}})$
will evolve via equation (\ref{kin}) to its static solution (\ref{stat}) just
because the static solution can not evolve any further.\footnote{\footnotesize
\baselineskip0.3cm The situation is not that simple, however, since the kinetic
equation at $f \gg 1$ has many static solutions, and our system first tend to
evolve to the turbulent distribution, see Appendix \ref{ssec:zero}, and would
stay there (which it does in our numerical experiments) if the phase transition
with a nucleation of the condensate would not occur.}
This implies that simple and straightforward numerical iteration procedure
has to be robust. Given the distribution function at time $t$ we calculate
the collision integral and then we advance the distribution function in time as
$f({{\bf p}},t+\Delta t) = f({{\bf p}},t) +
I_{\text{coll}}({{\bf p}},t) \Delta t$.

The evolution of the initial distribution function which
allows for the condensate formation consists of two stages.
In the first stage condensate is absent and the distribution function
is finite. In the second stage $n_c \neq 0$ and the distribution function
has $\delta$-function singularity. Consequently, we have to derive
the kinetic equation in the form suitable for the numerical integration
in each of those cases separately.

\subsection{Zero initial condensate}
\label{ssec:zc}
\begin{figure}
\psfig{file=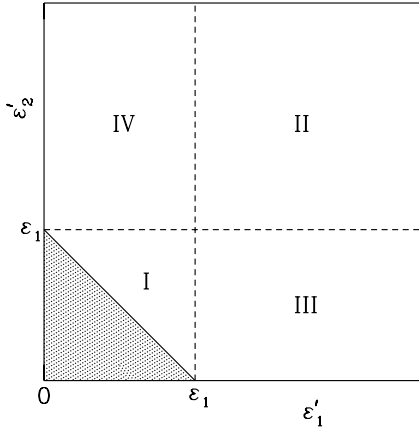,height=6cm,width=6cm}

\caption{The white areas correspond to the integration region
in the collision integral, Eq. (2.6)}.
\label{fig2}
\end{figure}

In  what follows we shall consider an isotropic initial distribution
$f=f(\varepsilon)$ and the non-relativistic limit
$\varepsilon={{\bf p}}^2/2m $.
The kinetic equation for the case without condensate, $n_c = 0$,
can be rewritten in the form (see Appendix \ref{sec:Simp}):

\begin{equation}
\frac{df (\varepsilon_1)}{dt} = \frac{|M_{fi}|^2}{64\pi^3m} \int \int
  F(f) \frac{D}{p_1} d\varepsilon'_1 d\varepsilon'_2  \equiv I_{\rm P}~,
\label{kin2}
\end{equation}
where
\begin{equation}
D \equiv \text{min} [p_1,p_2,p'_1,p'_2]~.
\label{min}
\end{equation}
The integration should be done over the white area of Fig. \ref{fig2} which is
divided into four regions.
In each of those regions we have
\begin{center}
\begin{tabular}{|cc|cc|} \hline
\rule{0mm}{4mm}
{}~$D = p_2$ & in I region~ & ~$D = p_1$ & in II region~ \\[1mm]
\hline
\rule{0mm}{4mm}
{}~$D = p'_2$~ & ~~in III region ~&~ $D = p'_1$~ & ~~in IV region~ \\[1mm]
\hline
\end{tabular}
\end{center}
In the argument of $F(f)$ according to the conservation of energy
$\varepsilon_2 = \varepsilon'_1+\varepsilon'_2-\varepsilon_1 $.

It is possible \cite{za66}, to map regions II-IV in Fig. \ref{fig2} to the
region I , and then to rewrite the collision integral in Eq. (\ref{kin2}) as an
integral over the
region I only. This representation is essential in finding the stationary
solutions to Eq. (\ref{kin2}) \cite{za66} and the corresponding Kolmogorov's
spectra, see Appendix \ref{self}.

\subsection{Non-zero condensate}
\label{ssec:sys}

This case we shall consider for bosons with self-interaction only.
After the moment of condensate  formation the kinetic equation (\ref{kin2})
is inappropriate for numerical integration, and  the finite number of
particles in the condensate corresponds to the infinite value of the
distribution function at zero energy.
In order to describe the system of particles interacting with the condensate
we divide the distribution function into two pieces:
\begin{equation}
\tilde{f} = f(\varepsilon,t) + (2\pi)^3 n_c(t) \delta^3 ({{\bf p}}) ~,
\label{fcon}
\end{equation}
where the first term corresponds to the "gas" of particles and the
second one describes the condensate.
Substituting this function into the original kinetic equation (\ref{kin})
we obtain
\begin{mathletters}
\label{kin3}
\begin{equation}
\dot{n_c}(t) = \frac{\lambda^2n_c(t)}{64\pi^3m} \int_0^\infty \!
d\varepsilon'_1
d\varepsilon'_2 [f'_1f'_2-f_2(1\!+\!f'_1\!+\!f'_2)]~,
\label{equationa4}
\end{equation}
\begin{equation}
\dot{f}(\varepsilon_1) = I_P +\frac{n_c(t)\lambda^2}{32\pi m^2p_1} \left(
\int^{\varepsilon_1}_0
[f'_1f'_2-f_1(1\!+\!f'_1\!+\!f'_2)]d\varepsilon'_2
 +  2 \int_{\varepsilon_1}^\infty [f'_2(1\!+\!f_1\!+\!f_2)-f_1f_2]
d\varepsilon'_2  \right) ~.
\label{fdot4}
\end{equation}
\end{mathletters}
In Eq. (\ref{fdot4}) $\varepsilon'_1=\varepsilon_1-\varepsilon'_2$ and
$\varepsilon_2=\varepsilon'_2-\varepsilon_1$, while in Eq. (\ref{equationa4})
$\varepsilon_2=\varepsilon'_1+\varepsilon'_2$.

In general, after the condensate formation (and  at large particle densities
even before) the description in terms of quasiparticles rather then particles
is more appropriate. For example, the kinetic equation (\ref{kin3})
does not include the processes where one of the incoming and one of the
outgoing particles have zero momentum, $p'_2 = p_2 = 0$. This process does not
contribute to the collision integral directly; {\it i.e.}, it does not change
the distribution of particles over energies, but it does change the
effective particle mass.
However, in many cases those effects are insignificant and we still can work in
terms of particles. Quantitatively, the original particle picture is correct if
the potential energy contribution from the interaction with the condensate to
the effective particle mass squared, $\Delta m_{\text{eff}}^2 =
\lambda\phi_c^2/6 = \lambda n/m$, is smaller than the particle momentum
squared,
{\it i.e.} $\lambda n/m \ll p^2$.  In fact, the kinetic equations of Refs.
\cite{e84} which contains all quasiparticle effects do coincide in this limit
with our Eq. (\ref{kin3}) derived in a simple way. Since $n \sim m^3 (\Delta
\upsilon)^3f_0$, we obtain $\lambda f_0(\Delta \upsilon) \ll 1$ as a condition
of validity of our equations, where $\Delta \upsilon$ is the characteristic
velocity dispersion.  In the case of axion miniclusters, for example,  we have
\cite{kt93} $\lambda f_0\sim 10^{-5}$; $\Delta \upsilon \sim 10^{-8}$, and the
description in terms of particles is perfectly good.

Note also that the kinetic equation itself becomes invalid at characteristic
energies smaller than the inverse relaxation time, but this condition is always
weaker than the previous one.

\section{Self-interacting bosons}
\label{sec:NC}
\subsection{Initial distribution and the rescaling of time}
\label{ssec:ic}
Let us first consider the system of self-interacting bosons.
As an initial distribution we choose the function $f(\varepsilon)$, which has
a maximum at ${{\bf p}}=0$. In general, such  a distribution function can
be characterized by means of three major parameters:

(1) The overall amplitude $f_0$. In what follows we define
$f_0 \equiv f(\varepsilon=0)$.

(2) The energy scale $\varepsilon_0$, where the distribution function
becomes 2 times smaller, $f(\varepsilon_0)=f_0/2$.

(3) The effective width $\Gamma$ of the region over which the distribution
function varies rapidly.

More specifically, we choose the initial distribution function to be of the
form:
\begin{equation}
f(\varepsilon)=\frac{2 f_0}{\pi}\arctan{\left[
\exp{(\Gamma(1-\varepsilon/\varepsilon_0))}\right]} ~.
\label{fin}
\end{equation}
In what follows we shall measure the distribution function in units of $f_0$,
{\it i.e.} the initial distribution function which will appear throughout the
rest of this
section will have the normalization $f(\varepsilon=0)=1$, and we shall
measure the energy in units of $\varepsilon_0$.

 We define the dimensionless time $\tau$ as \cite{it91}
\begin{equation}
\tau = \frac{\varepsilon_0^2 f_0 (1 + f_0) \lambda^2 }{64 \pi^3 m }~ t~.
\label{time}
\end{equation}

The parameter  $\varepsilon_0$ after rescaling will not enter the kinetic
equation explicitly, but will define the time scale of the evolution.
The parameter $f_0$ defines the time scale as well, but  it remains present in
the equation through the definition of F if $f_0 \sim 1$:
\begin{equation}
F(f)=\frac{f'_1f'_2 - f_1f_2 +
f_0 ([f_1+f_2]f'_1f'_2 - [f'_1+f'_2]f_1f_2 )}{1+f_0}.
\label{F2}
\end{equation}
Otherwise it also disappears from the rescaled equation: in the limit
$f_0 \ll 1$ one can neglect terms $\sim f^3$, and in the limit $f_0 \gg 1$
it is possible to drop terms $\sim f^2$ in Eq.(\ref{F2})

In terms of $\tau$ there remains the
weak dependence of the relaxation time upon the initial shape parameter
$\Gamma$. All data presented in this section will correspond to one and
the same value of $\Gamma = 5$.

After rescaling everything in Eq.(\ref{kin2}) is of order unity and,
very roughly, it is expected that the time-scale of the relaxation  towards
the equilibrium corresponds to $\tau \sim 1$, see Ref.\cite{it91}.

While the  distribution function Eq.(\ref{stat}) has three characteristic
parameters $\{ T,\mu, n_c \}$, only two of them are non-zero simultaneously.
Those two non-zero
parameters are uniquely defined by the values of two conserved quantities:
densities of the energy and the number of particles:
\begin{mathletters}
\label{trta}
\begin{eqnarray}
e &=& \int \varepsilon f({{\bf p}}) \frac{d^3{{\bf p}}}{(2\pi)^3}
\,\, ,
\label{equationa} \\
n &=& \int f({{\bf p}}) \frac{d^3{{\bf p}}}{(2\pi)^3}
\,\, .~~~~~~~~~~~
\label{equationb}
\end{eqnarray}
\end{mathletters}

\begin{figure}
\psfig{file=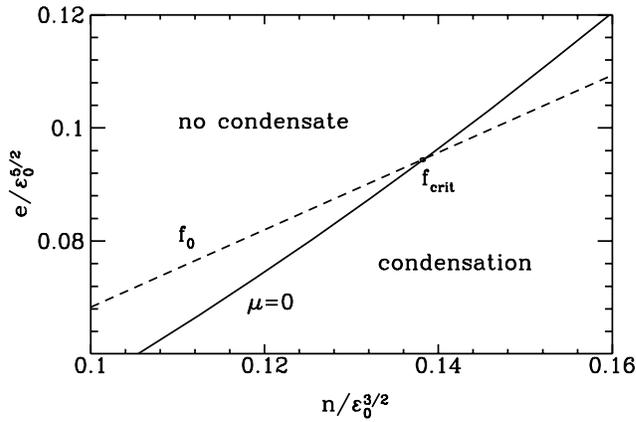,height=6cm,width=9cm}

\caption{The phase plane of the system. Below the solid line
$\mu = 0$ the condensate is non-zero in equilibrium. The dashed line
corresponds to the variation of the parameter $f_0$.}
\label{fig3}
\end{figure}

With the shape of the initial distribution function being given, the  two
scaling parameters $f_0$ and $\varepsilon_0$ in addition to fixing the time
scale define also the parameters of final equilibrium $\{ T,\mu, n_c \}$. This
correspondence is illustrated in Fig. \ref{fig3}.

The solid  curve divide the plane of Fig. \ref{fig3} in two regions and
corresponds to the equilibrium distribution function with $\mu = n_c = 0$.
Below this
curve the chemical potential $\mu$ remains zero, but $n_c>0$. {\it I.e.}, if
the initial number density of particles and the energy density
correspond to this  region, then the condensate forms in the final state.
On the contrary, for the region above the solid curve, $\mu < 0$ and
$n_c = 0$ in Eq.(\ref{stat}).
The dashed line on the Fig. \ref{fig3} corresponds to the variation of the
parameter
$f_0$ (keeping the parameter  $\Gamma$ of the initial distribution (\ref{fin})
to be fixed, $\Gamma =5$). With the increase of $f_0$ the phase point moves
along the dashed line from left to right. The choice
$f_0 \lesssim f_{\text{crit}} \approx 2.8$ corresponds to a Bose-gas density
less than critical, while $f_0 > f_{\text{crit}}$ corresponds to condensate
formation in the final equilibrium state.

\subsection{Kinetics without condensate formation in the equilibrium state}
\label{ssec:clas}

For completeness (and as a particular check of our numerical procedures) let us
first consider the case $f_0 \lesssim f_{\text{crit}}$.
We integrated kinetic equation with two particular values of $f_0$,
$f_0=0.1$ and $f_0=1$. In the case $f_0=0.1$ the $f^3$ quantum
terms in the function $F(f)$, Eq.(\ref{F2}), are small corrections and the
kinetic equation reduces to the Boltzmann equation for classical particles,
while $f_0=1$ is a true quantum regime.

\begin{figure}
\psfig{file=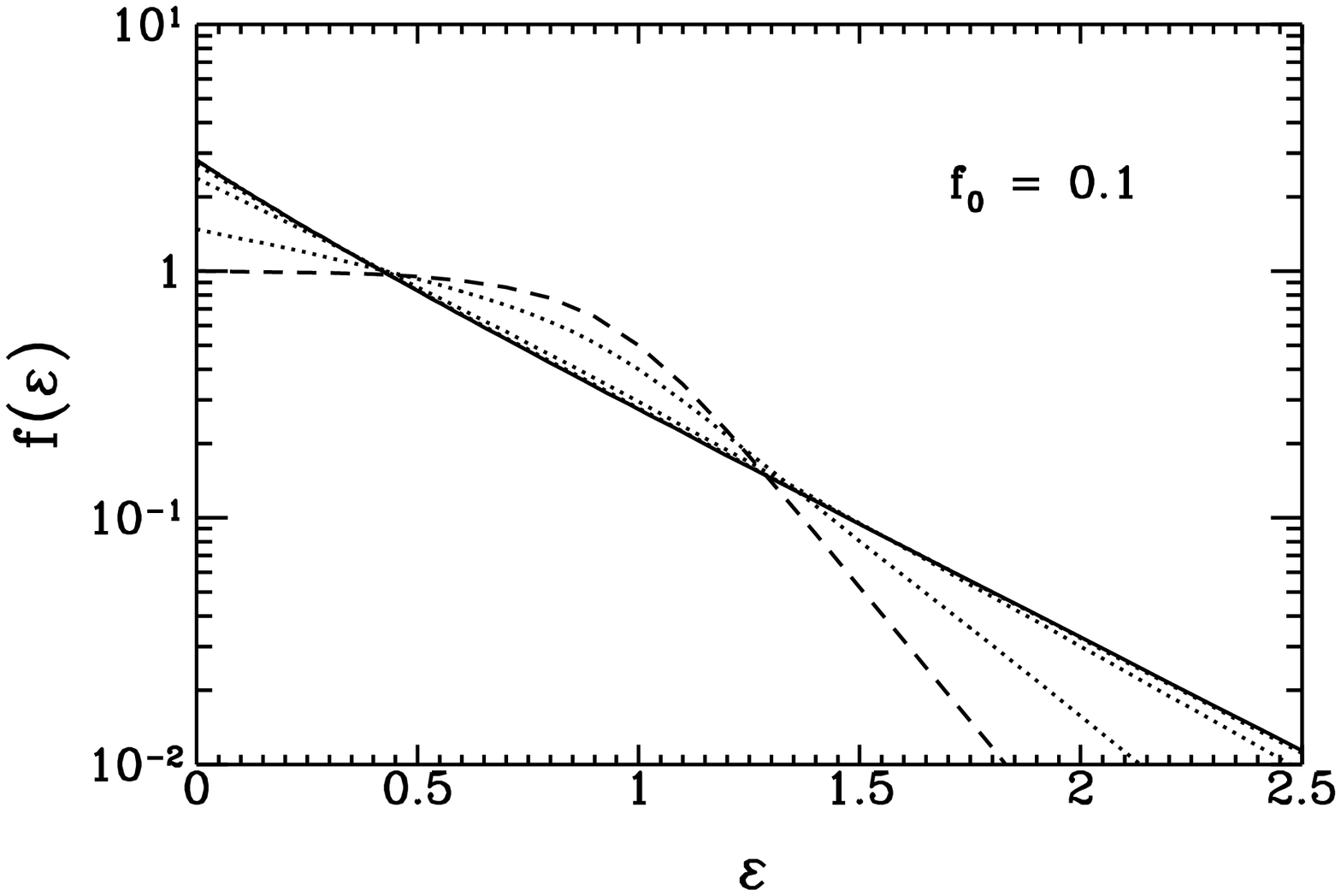,height=6cm,width=9cm}

\psfig{file=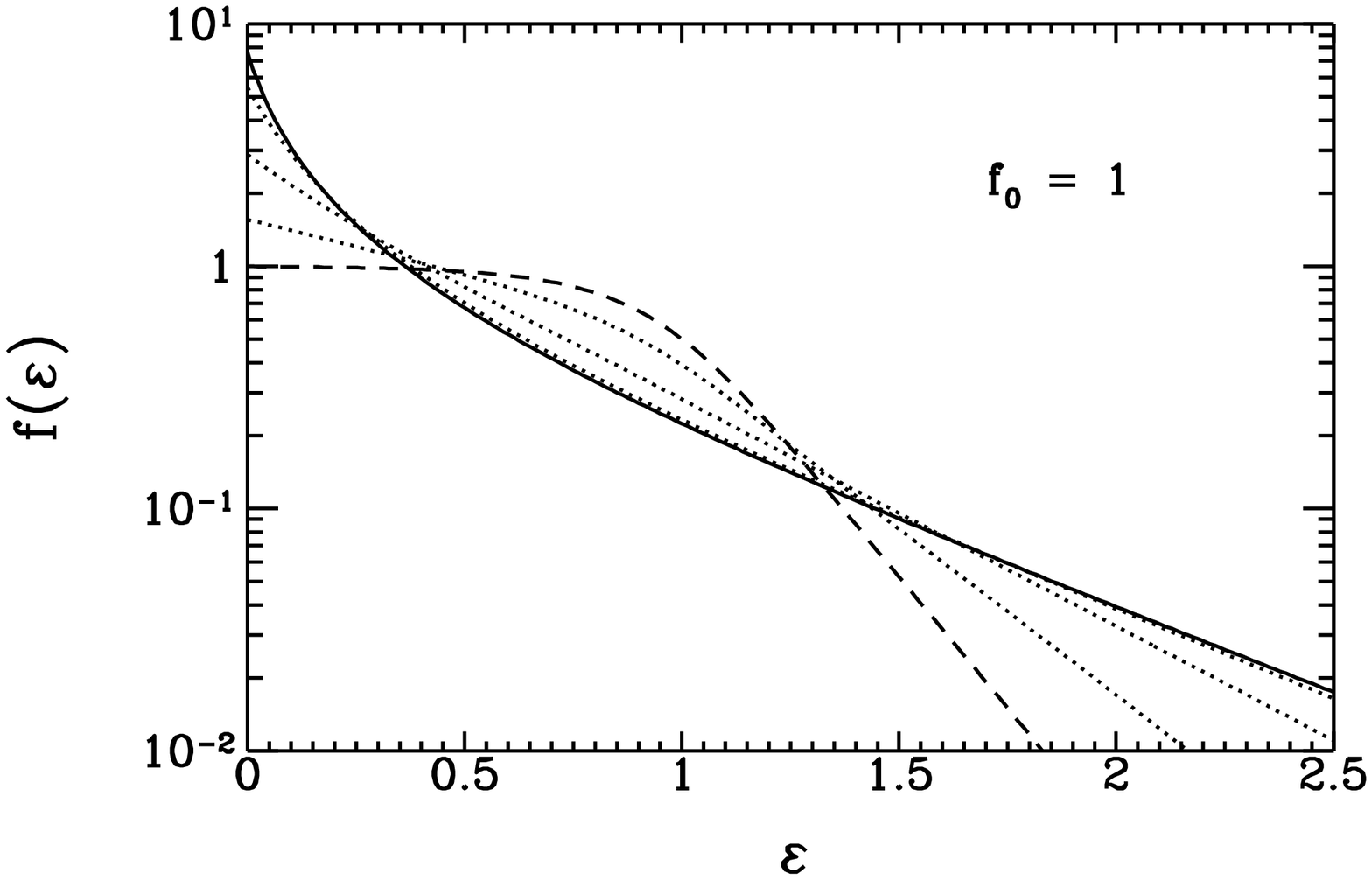,height=6cm,width=9cm}

\caption{Snapshots of the distribution function in non-condensate case. The
dashed line corresponds to the initial distribution; the equilibrium
distribution is presented by the solid curve;
dotted curves correspond to the results of numerical integration at
different moments of time $\tau = 1,5,10,20$.}
\label{fig4}
\end{figure}

The results of numerical integration of the kinetic equation
are presented in Fig. \ref{fig4}. The dashed
curve corresponds to the initial distribution function (\ref{fin}).
The dotted curves  represent numerical results at a sequence of
moments $\tau \approx 1, 5, 10, 20$. The equilibrium distribution
is shown by the solid curve. It is close to the classical Boltzmann
distribution in Fig. \ref{fig4}a, and corresponds to the Bose-Einstein
distribution
(\ref{stat}) with $T \approx 0.63\, \varepsilon_0$ and $\mu \approx
-0.078\, \varepsilon_0$ in Fig. \ref{fig4}b.
We see that already at $\tau \approx 10$ the numerical
results are almost indistinguishable from the equilibrium distributions,
and by the time $\tau \approx 20$ they coincide identically
on the scale of Fig. \ref{fig4}.

\begin{figure}
\psfig{file=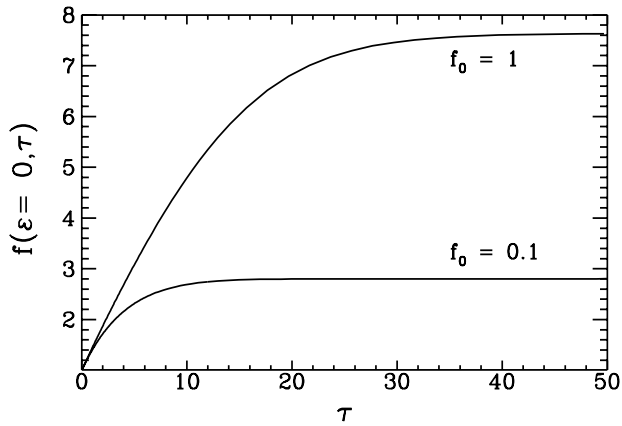,height=6cm,width=9cm}

\caption{The distribution function at zero energy as a function of time for two
cases presented in Fig. 4.}
\label{fig5}
\end{figure}

On Fig. \ref{fig5} the value of the distribution function at ${\bf p} =0$ is
shown as a function of time for both cases $f_0=0.1$ and $f_0=1$. We see that
the relaxation time is  close to 10 in the first case, while $\tau_{\text{rel}}
\approx 20$ in the second case.
However, according to Eq.(\ref{time}) the physical relaxation time in the case
$f_0=1$ is an order of magnitude shorter, which is the effect of the stimulated
scattering.

\subsection{Non-linear quantum kinetic regime or the classical turbulence}
\label{ssec:quan}
Now we turn to the case we are interested in: the Bose-condensate formation.
The initial distribution have to correspond to a large values of the parameter
$f_0$, {\it e. g.} in
our case of Eq. (\ref{fin}) $f_0 > 2.8$. We shall simplify the problem and
consider $f_0 \gg 1$. In this case
we can  disregard $f^2$ terms in the function (\ref{Fun1}),
which becomes:
\begin{equation}
F(f)=[f_1+f_2]f'_1f'_2 - [f'_1+f'_2]f_1f_2  ~.
\label{Ff}
\end{equation}
Parameter $f_0$ disappears from the kinetic equation and the latter became
scale invariant. This scale invariance gives the origin to a self-similar
evolution, as we shall see shortly.

As we have noted in the Introduction, very large occupation numbers corresponds
to the statistics of classical waves. The kinetic equation with $F(f)$ given by
Eq. (\ref{Ff}) can be obtained directly in a random phase approximation for the
classical waves and is well known in the theory of turbulence \cite{my,zmr85}.

We integrated the kinetic equation in the energy interval
$10^{-9} < \varepsilon < 10$. We had defined the distribution function on the
grid of 200 points equally spaced in the logarithm of energy and used the
spline interpolation when calculating the distribution function
at intermediate points. We have checked that the grid of 400 points produces
essentially identical results. For each integration in collision integral
we had implied  Gauss algorithm. Particle and energy non-conservation was of
order $10^{-3}$ for the entire time of integration.

The results of the numerical integration of the kinetic equation (\ref{kin2})
are presented in Fig. \ref{fig6}, where we plot the distribution function at
different moments of time. We arranged the output each time that
$f(\varepsilon = \varepsilon_{\text{min}},t)$ had increased by one order
of magnitude. The most striking feature of this plot is the self-similar
character of the evolution. The distribution function has a "core" where
$f(\varepsilon) \approx \text{const}$ and the radius of the core decreases with
time while the value of $f(\varepsilon)$ in
the origin grows. Self-similar solutions exhibiting this kind of behavior can
be parametrized as
$f(\varepsilon,\tau) =A^{-\beta}(\tau) f_s(\varepsilon/A(\tau))$,
where it is assumed $\beta > 0$ and $A(\tau) \rightarrow 0$ with the increase
of time. Outside the core the distribution function does not depend upon time
and is the power law to very good accuracy.

The value of the logarithmic derivative of $f(\varepsilon)$ is plotted in Fig.
\ref{fig6}b.
The dotted lines in Figs. 6 correspond to the limiting value
$\alpha = 7/6$ (see Appendix \ref{self}) but this value is never reached
prior to the moment of condensate formation. After that moment the character of
the evolution completely changes (see the next Section). Rather, with the
boundary condition $df/d\varepsilon =0$ at $\varepsilon =0$ the power law on
the tail is $\alpha =1.24$.

We believe that the difference of the exponent in our self-similar distribution
from the limiting critical exponent $\alpha =7/6$ is not a kind of the
numerical artifact. In favour of that we have the following arguments:

1. Doubling the number of grid points and/or integrator accuracy does not
change the distribution function appreciably. We have tried also different
integration algorithms and different boundary conditions at large energies.

2. We had observed that after the particle flux wave reaches the bottom
boundary of the integration interval $\varepsilon_{\rm min} =10^{-9}$ (which
happens at $\tau \approx 19$ in our case), the distribution rapidly  relaxes to
the strict power law $\varepsilon^{-7/6}$ throughout the whole integration
interval. This confirms, first, that our integration procedure correctly finds
the roots of the equation $I_{\rm coll}=0$, and, second, that the critical
exponent $\alpha =7/6$ can be achieved with the fitting boundary conditions
only, while in our case of condensate formation $f'(0)=0$. This can explain
also why the final distribution observed in Ref. \cite{bs91} was not resolved
from $\alpha=7/6$. Indeed, the minimum value of the energy on the grid in Ref.
\cite{bs91} was only $\varepsilon_{\rm min} = 10^{-3}$, while we have
$\varepsilon_{\rm min} = 10^{-9}$, so the results of Ref. \cite{bs91} are more
effected by the boundary conditions at small $\varepsilon$ and this happens in
a shorter time.

3. From the analytical point of view the $7/6$ critical exponent corresponds to
a stationary distribution, {\it i.e.} to a solution of the equation $I_{\rm
coll} =0$, while the shape of the self-similar distribution is defined by Eq.
(\ref{sysb}) below, with yet to be determined parameter $\beta$ which does not
have to coincide with $7/6$.

The time dependence of the distribution function
at $\varepsilon =0 $  is shown in Fig. \ref{fig7} by the solid line.
Vertical dot-dashed line corresponds to $\tau = \tau_c$ and
the distribution function asymptotically tend to
this line according to the law $f(0,\tau)  \propto (\tau_c - \tau)^{-2.6}$,
which is shown by the dotted line on this figure.
It is possible to find this time dependence at $\varepsilon = 0$
analytically, using the self-similarity  of the solution, see Section
\ref{sec:ss}.

\subsection{Condensation in the presence of condensate}
\label{ssec:cond}

\begin{figure}
\psfig{file=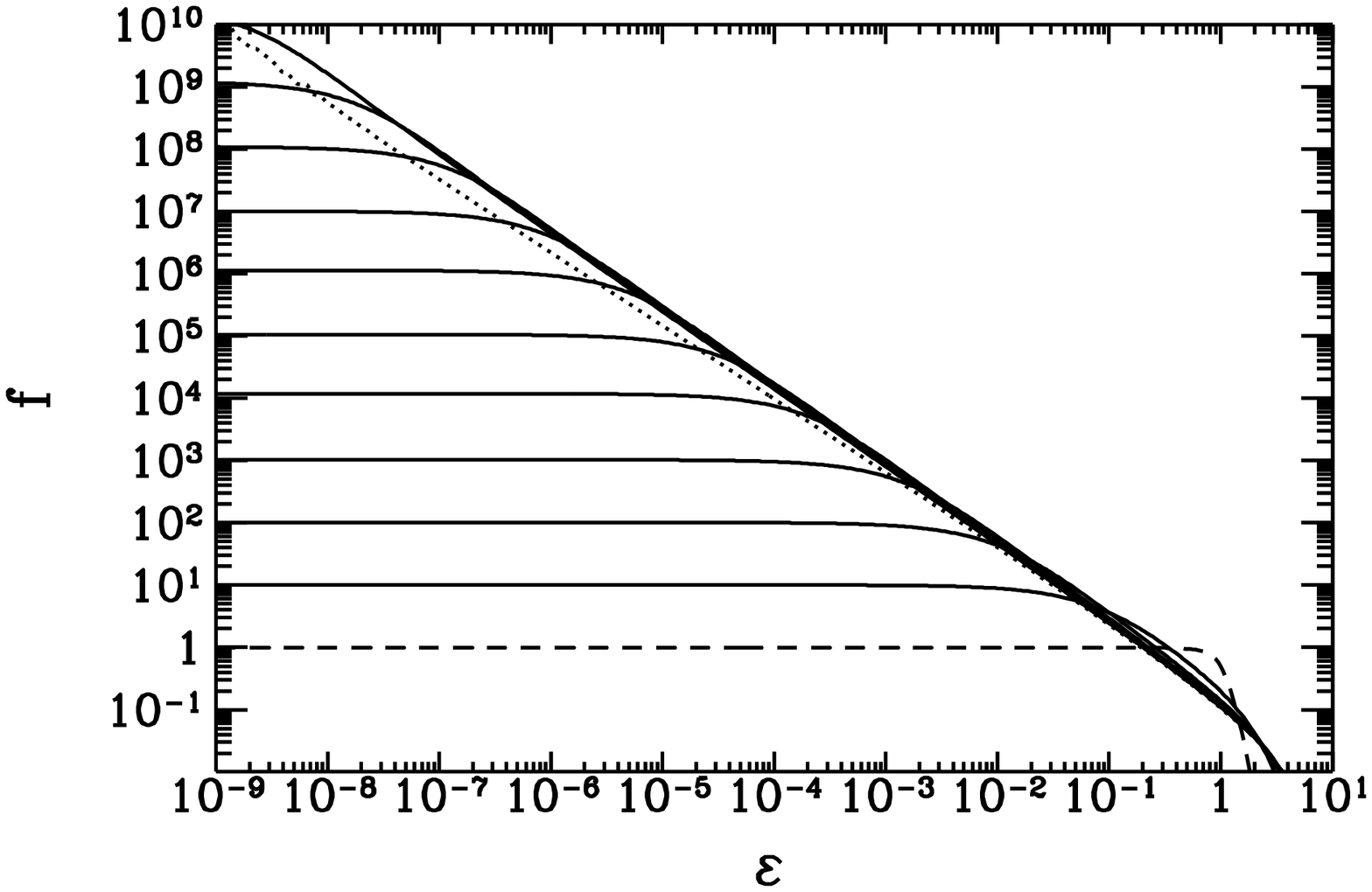,height=3.8in,width=6in}
\psfig{file=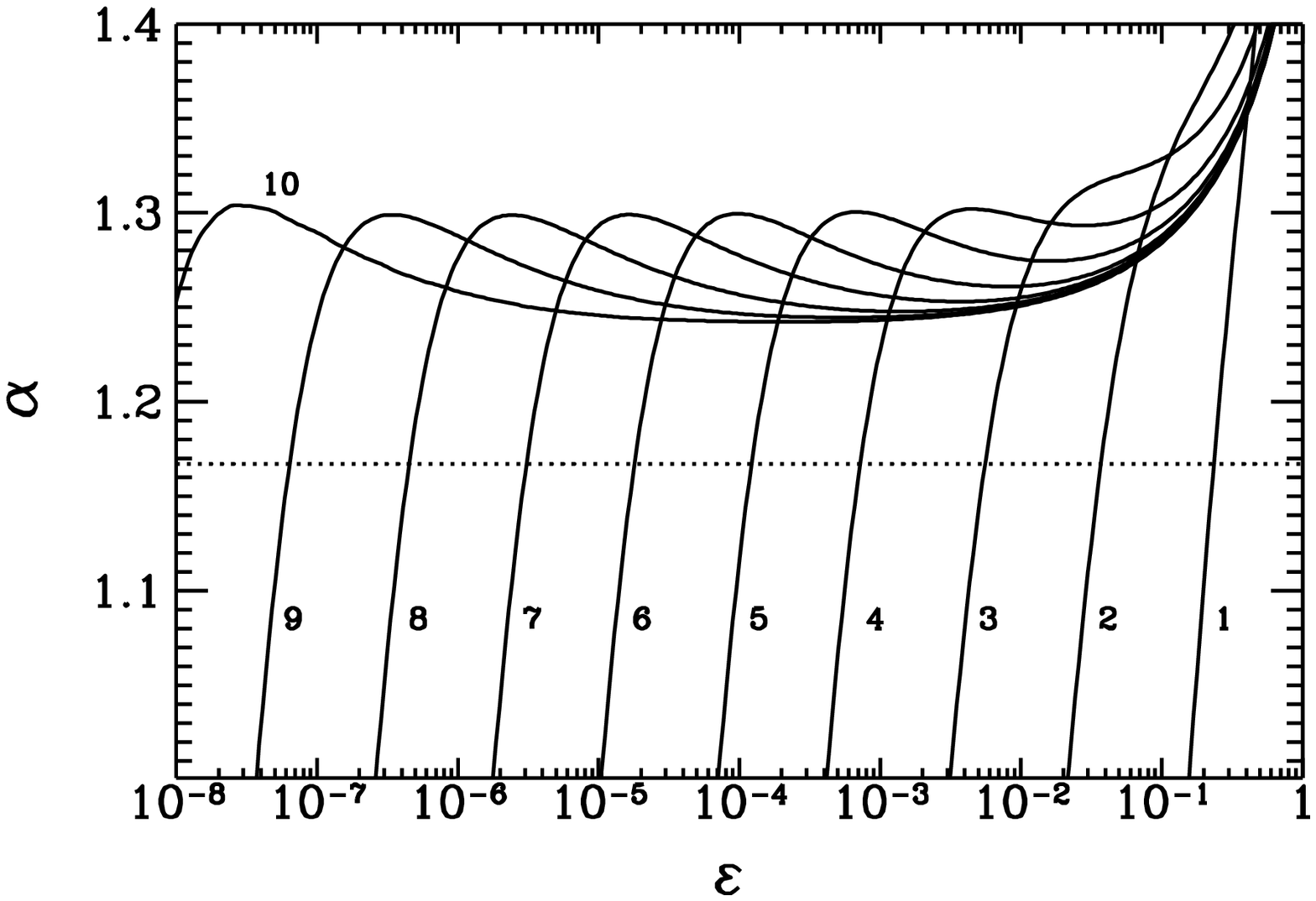,height=3.8in,width=6in}

\caption{The distribution function of self-interacting bosons (top) and its
logarithmic derivative (bottom) are shown at different moments of time prior to
condensate formation. The dashed line is the initial distribution; the dotted
line corresponds to the power law $f(\varepsilon) \propto \varepsilon^{-7/6}$.}
\label{fig6}
\end{figure}

In the previous subsection we have seen that the distribution function of
particles prior to the condensate formation tend to the power law with the
exponent $\alpha$ been almost constant, see Fig. \ref{fig6}. As one set of
initial conditions for the condensation in the presence of the condensate we
took the final distribution shown in Fig. \ref{fig6}. Though
the limiting value of $7/6$ was not reached prior to condensate formation, this
exponent is the special one and represents the root of the equation
$I_{\text{P}} = 0$, see Appendix \ref{ssec:zero}.
Because of that we have chosen the  function
\begin{equation}
f(\varepsilon)=C\, \varepsilon^{-7/6} e^{-\varepsilon}
\label{fincon}
\end{equation}
for the second set of the initial conditions while integrating
Eqs.(\ref{kin3}). In Eq. (\ref{fincon}) $C$ is some normalization constant, $C
= 1$ corresponds roughly to the magnitude of the final distribution at
$\varepsilon \sim \varepsilon_{\rm min}$ presented in Fig. \ref{fig6}a.
We took initially $n_c \ll n_{\text{tot}}$.
We will see shortly that the particular choice for $n_c$, as far as this
condition is satisfied, is insignificant.

First let us discuss the case with the initial condition (\ref{fincon}).
The results of the numerical integration of the system (\ref{kin3}) are
presented in Fig. \ref{fig8}. The dashed line  corresponds to the initial
distribution. Solid lines  correspond to the distribution
function $f(\varepsilon)$ at different moments of time. Basically, evolution
proceeds in the following way. First, the power law
$f(\varepsilon) \propto \varepsilon^{-7/6}$ changes to the law
 $f(\varepsilon) \propto 1/\varepsilon$ at small energies.
And then this change propagates to the region of larger energies,
see Figs. 9,10. Later on the power law stays at the equilibrium value
$\alpha = 1$, but the amplitude of the distribution function gradually
decreases. The same kind of  behavior of $f(\varepsilon,t)$ was found also in
the paper \cite{bs91}.

Again, we see that the essential part of the curves in Fig. \ref{fig8}b repeats
itself under translation from left to right and the evolution is self-similar.
During this epoch (before ``the wave of change'' had reached the exponential
tail of the initial distribution at $\varepsilon > 1$)
approximately $40\%$ of particles had condensed. And, what is important,
the number of particles in the condensate grows linearly with time at this
epoch, $n_c/n_{\rm tot}= B \tau$. We found $B \approx 2\, C^2$. This enables us
to eliminate the ambiguity in the initial value for $n_c$ since $B$ does not
depend upon it. Indeed, in our simulations which were done in a finite energy
interval, during the first several iterations the system self-adjusts: A proper
profile of the distribution forms while the
condensate reaches a particular value of $n_c$. We can disregard this period
and extrapolate the curves in Fig. \ref{fig8}b, $n_c(\tau )$, and the
self-similar character of the evolution back in time and to the region of
smaller energies.

The fraction of particles in condensate as a function of time is presented in
Fig. \ref{fig9}.

With the final distribution in Fig. \ref{fig6} taken as the initial condition
for the condensation in the presence of the condensate the fraction of
condensed particles increases with time dramatically differently initially.
Instead of $n_c/n_{\rm tot} \propto \tau$ we obtain now $n_c/n_{\rm tot}
\propto \sqrt{\tau}$. Both exponents can be understood analyzing self-similar
solutions, see Section \ref{sec:ss2}.

\section{Interaction with zero temperature fermion bath}
\label{sec:zeroT}

In this Section we consider ideal bose gas coupled to a fermion heat
bath of infinite capacity.  The function $F$ in the kinetic equation
(\ref{kin2}) in this case is defined by Eq. (\ref{Fun2}) and
$|M_{fi}|^2=g^4$.

For simplicity and definiteness we shall consider fermions at zero temperature,
$T_{\rm bath}=0$. The corresponding
distribution function of fermions has the form
$\eta(\varepsilon) =
\theta{(\mu-\varepsilon)}$, where $\mu$ is the Fermi energy. Such a simple form
of the fermion distribution
function allows us to make one more analytical integration in the collision
integral; at the end only one integration
remains to be calculated numerically. The price being paid for that is the
increased complexity of the final analytical expression which we put therefore
in Appendix \ref{zT}.

As an initial distribution we choose the function which has an equilibrium
form:
\begin{equation}
f_{\rm in}(\varepsilon)=\frac{1}{\exp{[(\varepsilon-\mu_{in})/T_{\rm in}]}-1}
{}~,
\label{finid}
\end{equation}
where $T_{\rm in}$ and $\mu_{\rm in}$ are initial temperature and chemical
potential of the Bose system correspondingly. In what follows we choose the
energy  scale
$\varepsilon_0$ and define the particle energy, the chemical potential and the
temperature in units of $\varepsilon_0$. To this end we choose $f_{\rm in}
(\varepsilon = 1) = 1$, which gives us
the relation $\mu_{\rm in} =[1- \log(2)] T_{\rm in} $. Under this condition
only one dimensionless parameter $T_{\rm in}$ is left to define the initial
state of Bose particles.

The parameter  $\varepsilon_0$ after rescaling will not enter
the kinetic equation explicitly, but will define the time scale of the
evolution. We define the dimensionless time $\tau$ as
\begin{equation}
\tau = \frac{\varepsilon_0^2  g^4 }{64 \pi^3 m }~ t~.
\label{time1}
\end{equation}

We had specified the initial distribution Eq. (\ref{finid}) choosing
$T_{\rm in}=2$ . For the Fermi energy  we have used $\mu_F = 20$. Since $T_{\rm
bath}=0$, all Bose particles has to condense eventually.

We integrated kinetic equation in the energy interval
$10^{-8} < \varepsilon < 100$. We defined the distribution function on
the grid of 661 points equally spaced in the logarithm of energy and
used Spline interpolation when calculating distribution function at
intermediate points. We had implied Simpson algorithm on the grid of the same
size for the collision integral .
The non-conservation of particle number over the integration time $\tau \approx
20$ was $\delta n/n = 0.001 \%$.

The results of numerical integration of the kinetic equation
(\ref{eq1}) are presented on Fig. \ref{fig10}, where we plot the distribution
function at several different moments of time.
The distribution function behaves completely differently now as compared to the
case of self-interacting bosons. It narrows in time
gradually approaching a $\delta$- function \cite{ly77}.

The value of the distribution function at zero energy (see Fig. \ref{fig11})
tend to infinity with time only as $\tau^3$.
Consequently, it would require infinite time for the
distribution function to reach the infinity. This does not necessarily means
that the condensation
time is infinite in this case (see the corresponding discussion in the
Introduction), but means that in this simple approach we can not find the
condensation time-scale for the ideal Bose gas coupled to
the thermal bath. However, we still can find the upper limit on the
condensation time-scale, even without the knowledge when the phase transition
to the condensed phase takes place actually. Indeed, the interaction with
fermions necessarily leads to the self-interaction in higher orders of the
perturbation theory (consider the box
diagram constructed from the diagram Fig. \ref{fig1}b) with $\lambda \sim g^4$.
When the product $g^4 f(0)$ exceeds O(1), this induced self-interaction starts
to dominate, and the problem reduces to the previous case of self-interacting
bosons with the explosive grows of the distribution function.

\section{Self-similarity and analytical solutions}
\label{sec:ss}

The results of the numerical integration presented in previous sections suggest
that
during condensate formation the distribution function evolve in a self-similar
way. Assuming self-similarity from the very beginning we can simplify the
kinetic equation and then a number of analytical results can be obtained
\cite{bs91,fs91,zlf}. In this
section we discuss this analytical approach and compare it to the numerical
results.

We parametrize self-similar solutions  as
\begin{equation}
f(\varepsilon,\tau) =A^{-\beta}(\tau) f_s(\tilde{\varepsilon})~,
\label{fst2}
\end{equation}
where $\tilde{\varepsilon} \equiv \varepsilon/A(\tau)$ and $\beta$ is some
constant. For definiteness we
shall assume $\beta >0$, {\it i.e.} in the following subsections {\bf A} and
{\bf C}, where we shall consider the epoch prior to the
condensate formation (when $f(0,\tau)$ has to grow), $A(\tau) \rightarrow 0$ as
time $\tau$ increases , while $A(\tau)$ will be growing function of time in
subsection {\bf B} which describes the epoch after the condensate has formed.
We always can choose the normalization $f_s(0) = 1$ if $f_s(0)$ is finite,
which is the case in subsections {\bf A} and {\bf C}, and $A(0)=1$ otherwise.

Self-similar solutions appear when the phase-space density is large, $f+1
\approx f$, in which case the kinetic equation became invariant with respect to
the rescaling. This
condition is satisfied around the condensate formation, and as we have seen,
the system approaches self-similar solutions starting with arbitrary initial
conditions. We assume $f(\varepsilon) \gg 1$ everywhere below.

\subsection{Self-interacting bosons prior to condensate formation}

Substituting parametrization (\ref{fst2}) into kinetic equation (\ref{kin2}) we
obtain:
\begin{equation}
-A^{-\beta-1}\dot{A}~ [\beta f_s(\tilde{\varepsilon}) + \tilde{\varepsilon}
\frac{df_s(\tilde{\varepsilon})}{d\tilde{\varepsilon}}] =
\frac{I_{\rm P}[f_s(\tilde{\varepsilon})]}{A^{3\beta-2}} ~.
\label{fs}
\end{equation}
Separating variables we find
\begin{mathletters}
\label{sys}
\begin{eqnarray}
& &-A^{2\beta-3}\dot{A}=C_s ~,\\
\label{sysa}
& &\beta f_s(\tilde{\varepsilon}) + \tilde{\varepsilon}
\frac{df_s(\tilde{\varepsilon})}{d\tilde{\varepsilon}} =
\frac{I_{\rm P}[f_s(\tilde{\varepsilon})]}{C_s}  ~,
\label{sysb}
\end{eqnarray}
\end{mathletters}
\noindent
where $C_s$ is (positive) separation constant.
Integrating the first equation in  (\ref{sys}) we obtain:
\begin{equation}
A(\tau)=[2 C_s (\tau_c-\tau)(\beta-1)]^{1/2(\beta-1)}~,
\label{aaa}
\end{equation}
where $\tau_c$ is the integration constant apparently corresponding to the
moment of time of the condensate formation. Indeed, the distribution function
at zero momentum as a function
of time is $f(0,\tau)=A(\tau)^{-\beta}$,
and $f(0,\tau)$ has the pole at $\tau=\tau_c$ if $\beta > 1$.

\begin{figure}
\psfig{file=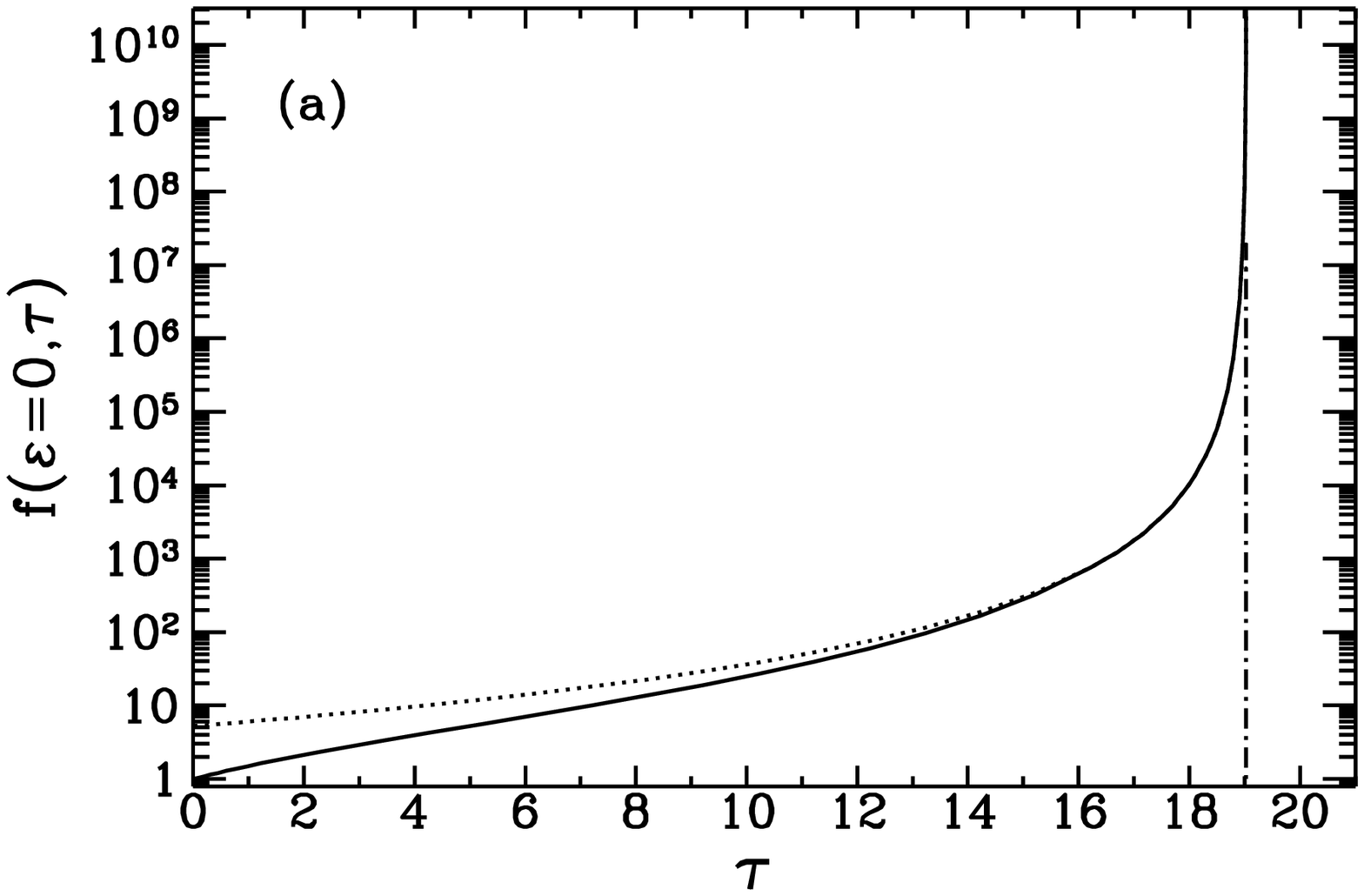,height=3.8in,width=6in}
\psfig{file=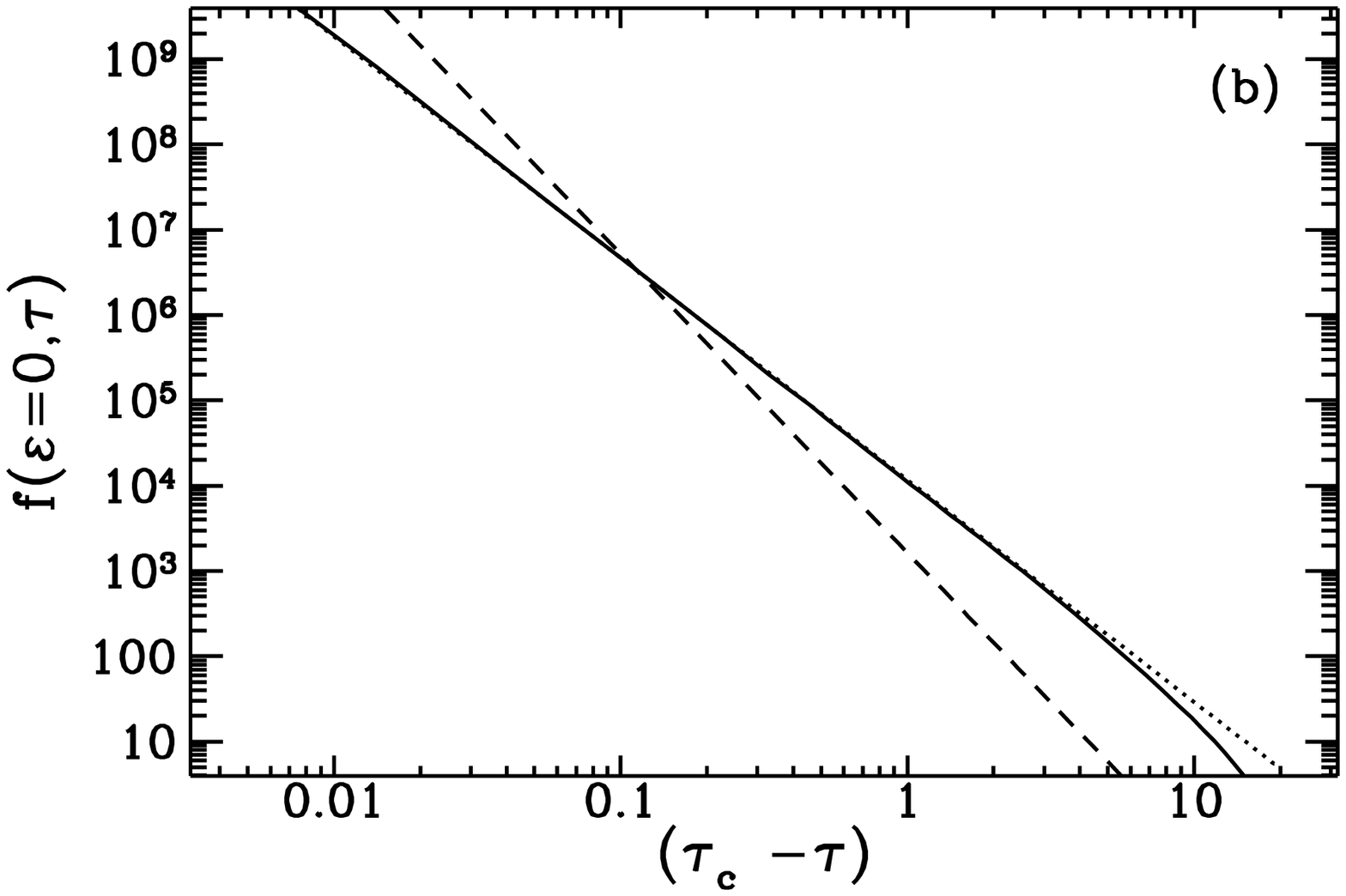,height=3.8in,width=6in}

\caption{The distribution function at $\varepsilon =0$ for the case presented
in Fig. 6 (the solid curves). The dotted curve in both figures is a solution to
self-similarity equations, $f \propto (\tau_c - \tau)^{-2.6}$, which was found
with $\alpha =1.24$, see Section V. The dashed line in Fig. (b) is the power
law $f \propto (\tau_c - \tau)^{-3.5}$ which would correspond to $\alpha
=7/6$.}
\label{fig7}
\end{figure}

\begin{figure}
\psfig{file=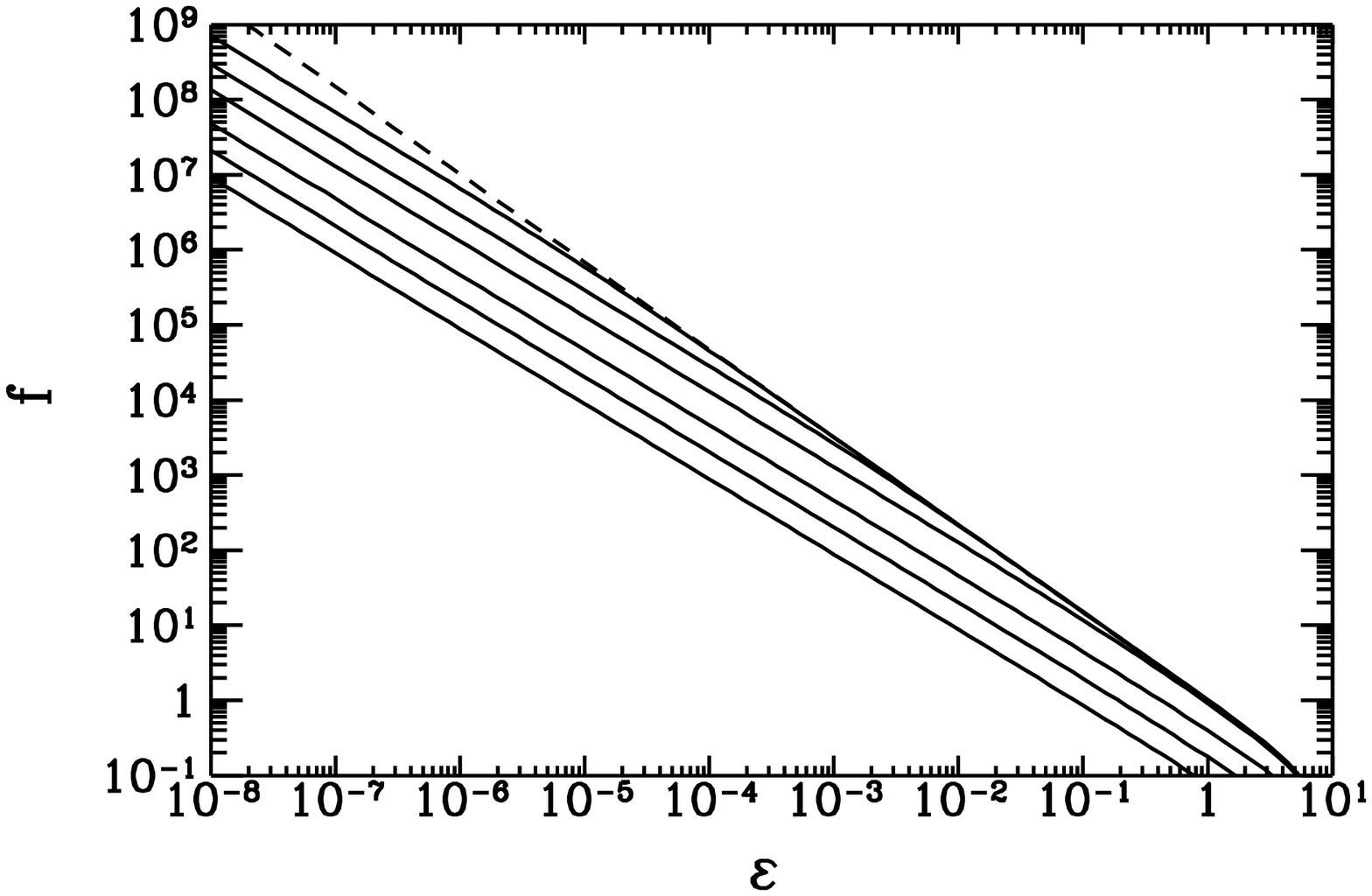,height=3.8in,width=6in}
\psfig{file=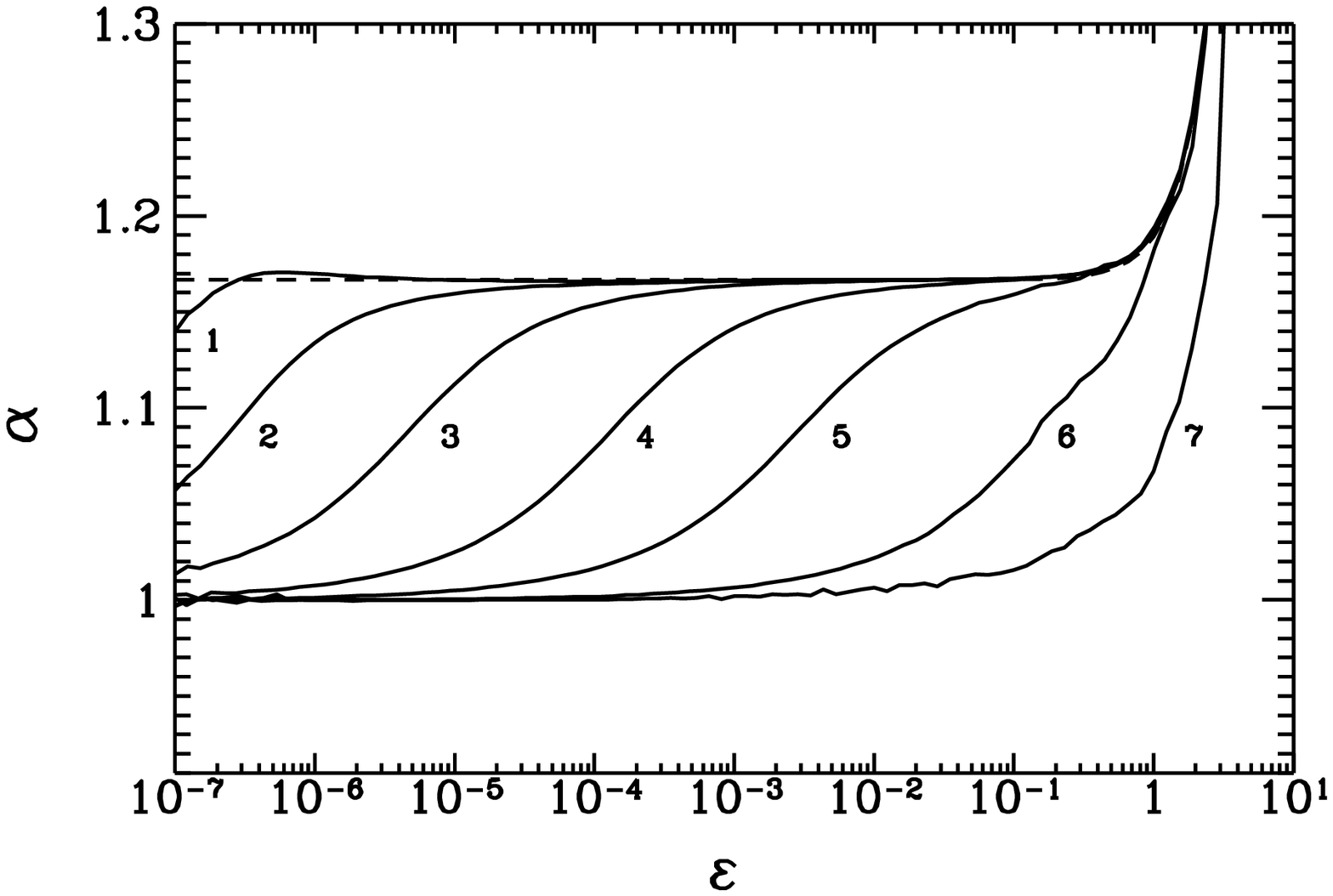,height=3.8in,width=6in}

\caption{ The distribution function of self-interacting bosons (top) and its
logarithmic derivative (bottom) are shown at different moments of time during
condensation. The dashed line is the initial distribution.}
\label{fig8}
\end{figure}

\begin{figure}
\psfig{file=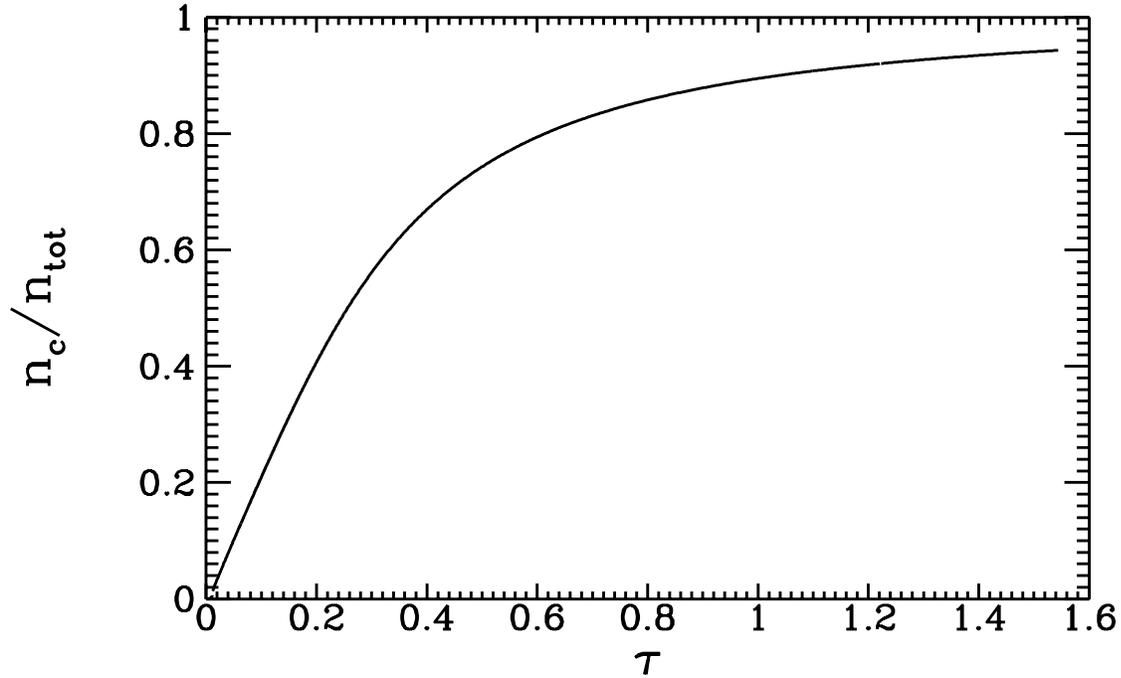,height=3.8in,width=6in}

\caption{Fraction of condensed particles as a function of time for the case
presented in Fig. 8.}
\label{fig9}
\end{figure}

We can conclude from Fig. \ref{fig6}a that at the tail where the distribution
function is a power law, $f \propto \varepsilon^{-\alpha}$, it does not
depend upon time. According to Eq. (\ref{fst2}) this means that
$\beta = \alpha$. This concludes the derivation of the function $f(0,\tau)$
for the self-similar solution, which is plotted in
Figs. \ref{fig7}(a,b) by the dotted line using $\alpha = 1.24$. This value of
$\alpha$ can be extracted from Fig. \ref{fig6}b . The agreement of this
function with direct numerical result is very good, see Fig. \ref{fig7}b. We
conclude, that after the solution have reached self-similar form, the time
dependence of the distribution function at zero momentum is given by
\begin{equation}
f(0,\tau) \propto (\tau_c - \tau)^{-\alpha/2(\alpha-1)} \propto
(\tau_c - \tau)^{-2.6}~,
\label{ftau}
\end{equation}
which reaches infinity at a finite time. Note that with the critical value
$\alpha =7/6$ we would obtain $f(0,\tau) \propto (\tau_c - \tau)^{-3.5}$. This
function is plotted by the dashed line in Fig. \ref{fig7}b and is quite
different from what we observe numerically.

\subsection{Self-interacting bosons and the condensate}
\label{sec:ss2}

Substituting parametrization Eq. (\ref{fst2}) into Eqs. (\ref{kin3}) we obtain:

\begin{mathletters}
\label{kin4}
\begin{equation}
\frac{d n_c(\tau)}{d\tau} = n_c(\tau) A(\tau)^{2-2\beta} I_c ~,
\label{ss1}
\end{equation}
\begin{equation}
\dot{A}(\tau) A(\tau)^{-\beta-1} \left[ \beta f_s({\tilde{\varepsilon}_1})
+\tilde{\varepsilon}_1 \frac{df_s({\tilde{\varepsilon}_1})}
{d\tilde{\varepsilon}_1}\right]
 = -A(\tau)^{2-3\beta}  I_P(\tilde{\varepsilon}_1)
- n_c(\tau) A(\tau)^{(1-4\beta)/2} I_{pc}(\tilde{\varepsilon}_1) ~,
\label{ss2}
\end{equation}
\end{mathletters}
where $I_c$ and $I_{pc}$ are integrals in r.g.s. of Eqs. (\ref{equationa4}) and
(\ref{fdot4}) respectively. Under the condition
\begin{equation}
\label{ncond}
n_c(\tau) = C_n A(\tau)^{(3-2\beta)/2}~~.
\end{equation}
variables in Eq. (\ref{ss2}) separates:
\begin{mathletters}
\label{kin5}
\begin{equation}
\dot{A}(\tau) A(\tau)^{-\beta-1} = A(\tau)^{2-3\beta} C_s ~,
\label{ss2a}
\end{equation}
\begin{equation}
-C_s \left[ \beta f_s(\tilde{\varepsilon}_1)
+ \tilde{\varepsilon}_1 \frac{df_s({\tilde{\varepsilon}_1})}
{d\tilde{\varepsilon}_1}\right]
 =   I_P(\tilde{\varepsilon}_1) + C_n I_{pc}(\tilde{\varepsilon}_1)~~.
\label{ss4a}
\end{equation}
\end{mathletters}
If we solve Eq. (\ref{ncond}) with respect to $A(\tau)$  and substitute it into
Eq.(\ref{ss1}) we find
\begin{equation}
n_c(\tau) = C_n\left[
\frac{4(1-\beta)I_c }{(2\beta-3)}\, \tau \right]^{(3-2\beta)/{4(\beta-1)}}
{}~,
\label{ss0a}
\end{equation}
which gives the time dependence of the number of condensed particles. As a
self-consistency check we find that the same substitution, but with Eq.
(\ref{ncond}) solved with respect to $n_c(\tau)$, gives the condition
equivalent to Eq. (\ref{ss2a}) if $C_s =2I_c/(3-2\beta)$. This defines $C_s$.
The solution of  Eq.(\ref{ss2a}) is:
\begin{equation}
A(\tau) = \left[ 1 +
\frac{4(1-\beta)I_c }{(2\beta-3)}\, \tau \right]^{{1}/{2(\beta-1)}}
{}~.
\label{ss0}
\end{equation}

As we have seen, the equilibrium profile $f \propto
1/\varepsilon$ propagates from the region of small energies to the region of
large energies as a ``wave'', leaving $f \propto
\varepsilon^{-\alpha}$ to be time independent in the region which this ``wave''
has not reached yet. This, again, means that $\beta = \alpha$. With
$\alpha=7/6$ as an initial condition we find from Eq. (\ref{ss0a}) that the
condensate linearly grows with time, $n_c(\tau) = {\rm const} \times \tau$.
This is in agreement with the results of our numerical experiment, see Fig.
\ref{fig9}. This self-similar regime persisted till the ``wave'' have reached
exponential tail where $\alpha$ is different. During this period $\approx 40\%$
of particles had condensed. With the initial condition $\alpha=1.24$ we obtain
dramatically different rate of condensate growth, $n_c(\tau) \propto
\tau^{0.54}$ which, again, was confirmed by the direct numerical integration.

\subsection{Interaction with degenerate Fermi gas}
\label{sec:ssbf}

\begin{figure}
\psfig{file=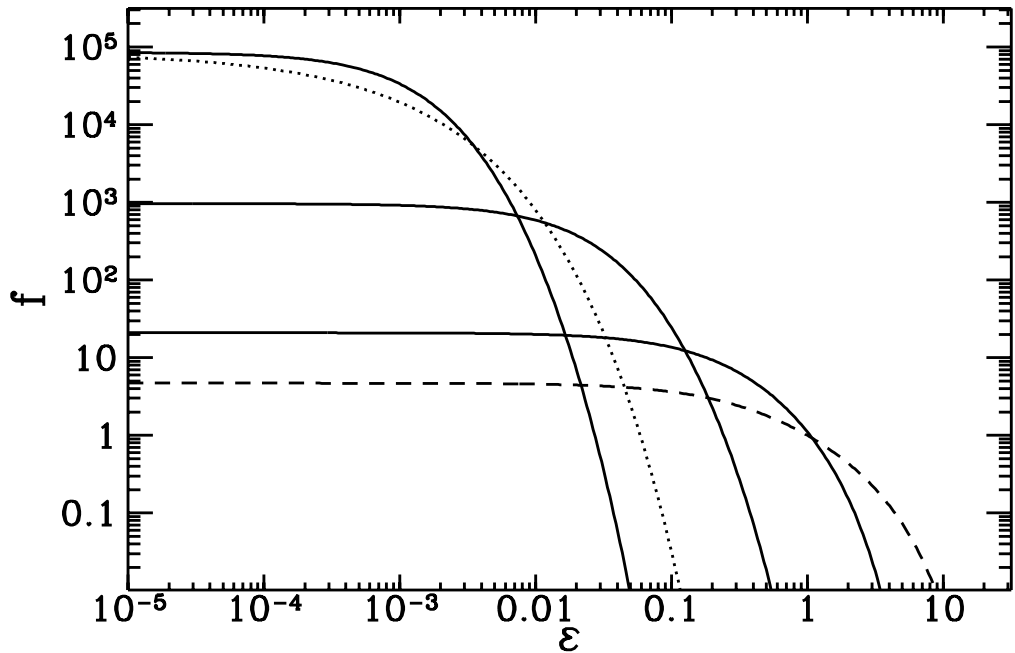,height=3.8in,width=6in}

\caption{The distribution function of bosons interacting with the cold gas of
fermions is shown at different moments of time by the solid curves. The dashed
curve is the initial distribution, the dotted curve corresponds to the
Fokker-Planck approximation of Ref. [13].}
\label{fig10}
\end{figure}

\begin{figure}
\psfig{file=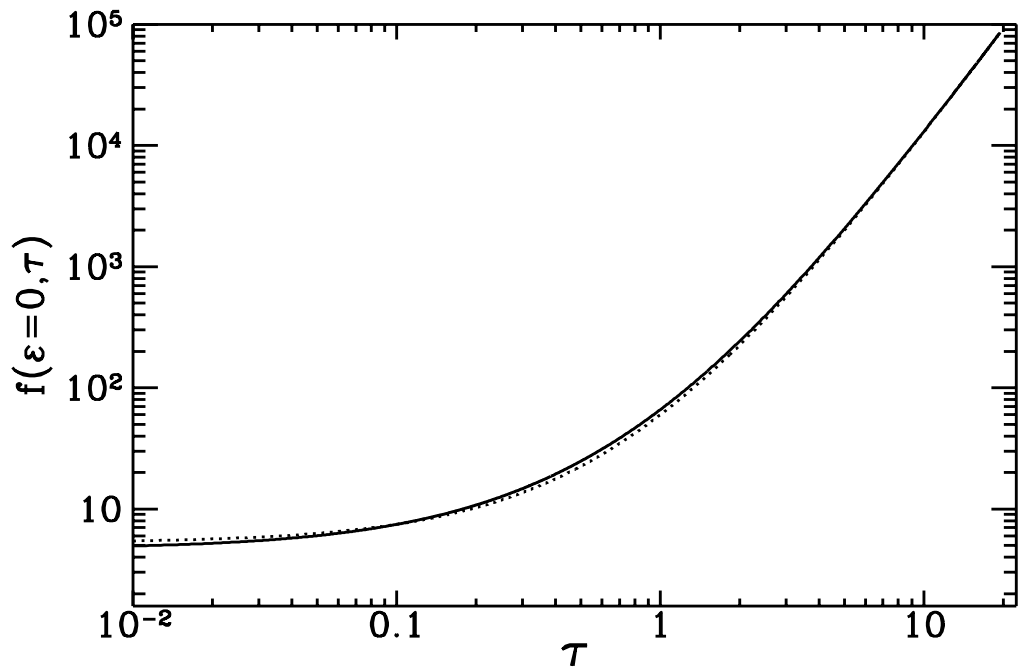,height=3.8in,width=6in}

\caption{The distribution function at $\varepsilon =0$ for
the case presented in Fig. 10. The dotted line is the power
low $f \propto (\tau+\tau_0)^3$, which is a solution to self-similarity
equations, see Section V.}
\label{fig11}
\end{figure}

Now we substitute parametrization (\ref{fst2}) into the kinetic equation with
the function F given by Eq. (\ref{Fun2}).
Left-hand side of equation (\ref{fs}) remains unchanged in this case, however,
its
right-hand side reads now $I[f_s(\tilde{\varepsilon})]/A^{2\beta-2}$. This
changes equation (\ref{sys}a) to
\begin{equation}
\dot{A} =-A^{3-\beta}\, C_s ~ .
\label{dA2}
\end{equation}
In the case of the
self-interacting bosons we were able to find $\beta$ using the information
gained from the numerical integration that on the tail of the self-similar
distribution the value of $f$ does not depend upon time.  That gave us
$\beta=\alpha$ and we have extracted $\alpha$ as a logarithmic derivative of
the distribution function with respect to the energy. We could not use
integrals of the motion in that case since the particle number
(as well as the energy density) saturated at the very end of the distribution
where the
self-similarity does not hold anymore (since on the self-similar solution
we had $\alpha < 3/2$). Now, on the contrary, the particle number saturates
already near the core of the self-similar distribution, see Fig. \ref{fig10}
(moreover, $\alpha$ grows with
time). This enables us to find $\beta$ using conservation of particles.

Substituting parametrization (\ref{fst2}) into Eq. (\ref{equationb}) we find
$\beta=3/2$. Consequently, the equation (\ref{dA2}) reads now $\dot{A} =
-A^{3/2}\,C_s$.
We find finally that
\begin{equation}
f(\varepsilon=0,\tau) = A^{-3/2} \propto (\tau+\tau_0)^3 ~ .
\label{fbath}
\end{equation}
This is in good agreement with the results of our numerical integration, see
Fig. \ref{fig11}.

The $\varepsilon$-dependence of the distribution function can be found from the
equation (\ref{sysb}), which in the limit of large $\mu$ takes the form
\begin{equation}
\frac{3f_s}{2}+\tilde{\varepsilon}\frac{df_s}{d\tilde{\varepsilon}}
= \frac{f_s}{C_s}\left[ \frac{1}{\sqrt{\tilde{\varepsilon}}}
\int_0^{\tilde{\varepsilon}}dxx^{1/2}(x-\tilde{\varepsilon})f_s(x)+
\int_{\tilde{\varepsilon}}^\infty dx (x-\tilde{\varepsilon})f_s(x) \right]
\label{fsn}
\end{equation}

Let us make here the connection to results of Ref. \cite{ly77}. One can see
that it is possible to rewrite the final expressions of \cite{ly77} in a
self-similar form, Eq. (\ref{fst2}), with the same value of $\beta =3/2$ and
the same function $A(\tau) \propto \tau^{-2}$ at large $\tau$. This gives $f_s
= f_s (\sqrt{\varepsilon} \tau)$. Using this and noting that Levich and Yakhot
started from the initial conditions corresponding to the critical point,
$f_s(\tilde{\varepsilon}) \propto 1/\tilde{\varepsilon}$, we find their result
$f(\varepsilon,\tau) \propto \tau$. Our direct numerical integrations with an
appropriate initial conditions confirm that. However, in the runs the results
of which we had reported above we have had different boundary conditions,
$f_s(0)=1$, which leads to $f(0,\tau) \propto \tau^3$ at large $\tau$, Eq.
(\ref{fbath}).  We see that the system which is father away from the critical
point condenses with a higher rate.

The solution found in Ref. \cite{ly77} for $f_s$, which with our initial and
boundary conditions corresponds to $f_s(\tilde{\varepsilon})=\exp (-\sigma
\sqrt{\tilde{\varepsilon}})$, is not an exact solution to equation (\ref{fsn}),
however, it is a reasonable approximation.
This solution is plotted in Fig. \ref{fig10} by the dotted line.

\section{Conclusions and discussion}
\label{sec:Con}
We have studied numerically the kinetics of condensation of
the weakly interacting Bose gas. We have found that the distribution function
evolves very differently in cases of Bose-bath interaction and
self-interaction of Bose particles. This results from the different structure
of the kinetic equations describing particle-bath and particle-particle
interactions [the latter contains extra powers of $f(\varepsilon)$]. In the
first case the distribution function of the excess particles, which eventually
form the condensate, narrows gradually with time and continuously
approaches a $\delta$- function in infinite time, as it was found in Ref.
\cite{ly77} assuming small energy exchange per collision.

For the case of self-interacting bosons we find, in qualitative agreement with
Refs. \cite{bs91,kss92}, that the singular power law profile $f(\varepsilon)
\propto \varepsilon^{-1.24}$ of the distribution function forms in a finite
time. This is close to $f(\varepsilon) \propto \varepsilon^{-7/6}$ which is a
stationary solution of the kinetic equation and corresponds to the constant
flux of particles in momentum space towards the condensate. Different exponents
result from different boundary conditions during the Bose condensation, namely
$\partial f/\partial \varepsilon (0,t) =0$, as opposed to the case of
stationary turbulence.  While the exponents differ by only 6\% this leads to
significantly different values for other critical exponents, {\it e.g.} $\beta$
in $f(0,t) \propto (t_c -t)^{-\beta}$ , or for the rate of the condensate
growth $\gamma$ in $n_c \propto t^\gamma$. We have found $\beta=2.6$ and
$\gamma =0.54$, while for $f(\varepsilon) \propto \varepsilon^{-7/6}$ we would
have $\beta=3.5$ and $\gamma =1$.

One could expect that this  very natural regime of the constant flow of
particles into the condensate will persist in the presence of the condensate as
well until all excess particles from the high energy tail are in the
ground state. Nevertheless, this is not the case and at the moment the
condensate appears, its presence terminates this regime. Instead of such a
steady flow through the entire energy interval, particles from all energy
levels jump directly to the condensate, while the remaining particles maintain
an equilibrium shape of the distribution function $f(\varepsilon) \propto
\varepsilon^{-1}$. The constant of proportionality in this law gradually
decreases until it reaches an equilibrium value.

The build up of coherence cannot be observed in the
framework of the Boltzmann kinetic equation, but the kinetic description has to
be valid prior to and after the moment of condensate formation. We have seen
that evolution at both stages is self-similar. This allows us to obtain a
number of useful analytical relations, {\it e.g.} the time dependence of the
distribution function near the point of condensate formation, Eq. (\ref{aaa}),
and Eq. (\ref{ss0a}) for the subsequent growth of the condensate. We have found
the duration of both stages, which is finite and of order $\tau_c \sim 20$.

Our results of numerical integration of the Boltzmann equation might be used
as an initial condition in a theory of condensate nucleation.

Another interesting application of our results could be in a description of
Bose-star formation. A complete theory of this process requires the
inclusion of gravity and the problem is more involved. While the isotropy in
momentum space will be broken, the kinetics of Bose relaxation in the
collapsing core can proceed largely along the lines described in the present
paper. Moreover, the relaxation time can become shorter, since the spatial
collapse of the core under the influence of self-gravity would cause an
additional increase of the density of particles, which would speed up the
relaxation.

It is remarkable that in spite of the apparent smallness of the axion quartic
self-coupling, $\lambda_a^4 \sim 10^{-53}
f_{12}^{-4}$ (where $f_{12} = f_a/10^{12}$ GeV)  the relaxation time in axion
miniclusters can be comparable to the age of the universe \cite{it91,kt93}.
This is because the relevant quantity is not the self-coupling itself, but the
product of $\lambda$ with the mean field strength squared (or with the mean
phase-space density of ``particles'', which in the present case would rather be
the phase-space density of classical scalar waves). For particles bound in a
gravitational well, it is convenient to rewrite the
expression for the relaxation time in the form
$\tau_c  \sim  m^7 \lambda^{-2}\rho ^{-2}v_e^2 $ \cite{it91}, where $\rho$ is
the energy density in a minicluster and $v_e$ is the escape velocity.
For axions, the relaxation time is smaller than the present age of the
Universe if the energy density satisfies
$\rho_{10}  >  10^{6} v_{-8}\sqrt{f_{12}}$,
where $\rho_{10} \equiv \rho /(10 \, {\rm eV})^4$ and $v_{-8}
\equiv v_e /10^{-8}$. It was shown in Refs. \cite{kt93} that those
large densities can be achieved naturally in axion miniclusters and, therefore,
a ``Bose-star'' can form \cite{gc}.

Our results are applicable whenever Eq. (\ref{kin}) holds and might also be
interesting for a study of, {\it e.g.}, Kolmogorov turbulence.

\acknowledgments
We thank V. A. Berezin, E. W. Kolb, V. A. Rubakov, G. Starkman and F. V.
Tkachov for useful discussions.
This work was supported by DOE and NASA grant NAG5-2788 at Fermilab.
D.S. thanks the Astrophysics Department of FNAL for the hospitality during
this work.

\appendix

\section{Reduction of the collision integral}
\label{sec:Simp}
In this Appendix we will make as many integrations in the collision integral
analytically, as it is possible. In the case we are interested in, {\it i.e.},
four-particle interaction vertex and the isotropic one-particle distribution
function, seven out of nine integrations can be done and only two integrals
upon the energies of incoming particles remain the for numerical treatment. We
start from the non-relativistic counterpart of Eq.(\ref{kin})
\begin{equation}
I_{\text{coll}} = \frac{(2\pi)^4}{16m^4} \int
\delta^4(\sum_i p_{\mu i}) |M_{fi}(|{\bf p} |)|^2  F(f) \prod_i
\frac{d^3{\bf p}_i}{(2\pi)^3 }~,
\label{Icoll}
\end{equation}
where $F(f)$  is defined in (\ref{Fun}) with the distribution function $f$
which depends upon particle energy $\varepsilon$ only. We shall make use of
the identity
\begin{equation}
\delta^3(\sum {\bf p}_i)=
\int  e^{i({\bf \lambda},{\bf p}_1+{\bf p}_2-
{\bf p}'_1-{\bf p}'_2)}\, \frac{d^3 {\bf \lambda}}{(2\pi)^3 }
\label{delta}
\end{equation}
and explicitly separate out angle integrations
\begin{equation}
\nonumber
d^3p_i = d\phi_i d\cos\theta_i p_i^2dp_i \equiv m p_i d\Omega_i
d\varepsilon_i~.
\end{equation}
Collision integral (\ref{Icoll}) takes the form:
\begin{equation}
I_{\text{coll}} = \frac{|M_{fi}|^2}{64\pi^3m} \int
\delta(\varepsilon_1+\varepsilon_2-\varepsilon'_1-\varepsilon'_2)  F(f)
D d\varepsilon'_1 d\varepsilon'_2 d\varepsilon_2 ~.
\label{kin1}
\end{equation}
where we have assumed that the matrix element does not depend upon momenta and
we have defined
\begin{equation}
D \equiv \frac{p_2 p'_1 p'_2}{64 \pi^5}\int \lambda^2 d\lambda
\int e^{i({\bf p}_1,
{\bf \lambda})}d\Omega_\lambda \int e^{i({\bf p}_2,
{\bf \lambda})}d\Omega_{p_2} \int e^{i({\bf p}'_1,
{\bf \lambda})}d\Omega_{p'_1} \int e^{i({\bf p}'_2,
{\bf \lambda})}d\Omega_{p'_2}~.
\label{V}
\end{equation}
All angle integrals are trivial in (\ref{V}) and we get
\begin{equation}
D = \frac{4}{\pi p_1}\int_0^\infty \frac{d\lambda}{\lambda^2}
\sin(\lambda p_1) \sin(\lambda p_2) \sin(\lambda p'_1)  \sin(\lambda p'_2)~.
\label{V11}
\end{equation}
After some algebra and taking into account the energy conservation condition
this gives
\begin{equation}
D = \frac{1}{p_1} \text{min} [p_1,p_2,p'_1,p'_2]~.
\label{V2}
\end{equation}
Now, the energy  $\delta$-function can be trivially integrated away and we are
left with  two-dimensional integral upon energies of incoming particles
\cite{za66}
\begin{equation}
\frac{df (\varepsilon_1)}{dt} = \frac{|M_{fi}|^2}{64\pi^3m} \int \int
  F(f) D d\varepsilon'_1 d\varepsilon'_2 ~,
\label{kin2a}
\end{equation}
where $\varepsilon_2 = \varepsilon'_1-\varepsilon'_2-\varepsilon_1 $.

\section{Stationary solutions to the kinetic equation}
\label{ssec:zero}

\subsection{Kolmogorov's spectra}
\label{self}

In this Appendix we shall describe power law solutions of the equation
$I_{\text{coll}} = 0$.
We shall consider the region of large values of the  distribution function,
$f(\varepsilon ) \gg 1$, {\it i.e.}, $F(f)$ will be given by Eq. (\ref{Ff}).
The kinetic equation with this form of $F(f)$ was extensively studied in the
literature devoted to the problem of the
Kolmogorov turbulence \cite{k41}, and in Ref. \cite{za66}
all power law solutions to the equation $I_{\rm coll} = 0$ were
found. Two of those solutions are evident. One is just the limit of the
Eq.(\ref{stat}) at $\varepsilon \rightarrow 0$, and $n_c =0 $ and has the form
$f = \text{const}/\varepsilon$. Another one is $f=\text{const}$.
Both are solutions to a simpler equation $F(f) \delta(\sum\varepsilon) = 0$.
To find the remaining solutions is a non-trivial task and with the distribution
function of the form $f(\varepsilon) = \varepsilon^{-\alpha}$
the trick is to map all labelled regions in Fig. \ref{fig2} to the region I by
means of Zakharov transformations \cite{za66}:

\begin{center}
\begin{tabular}{c|c|c}
$\text{II} \rightarrow \text{I}$ & $\text{III} \rightarrow \text{I}$ &
$\text{IV} \rightarrow \text{I}$ \\\hline
\rule{0mm}{3.5mm}
{}~$\varepsilon'_1 \rightarrow
\varepsilon_1^{} \varepsilon'_2/\varepsilon_2^{}$~ & ~$\varepsilon'_1
\rightarrow
\varepsilon_2^{}\varepsilon_1^{}/\varepsilon'_2$~ & ~$\varepsilon'_1
\rightarrow
\varepsilon_1^2/\varepsilon'_1$ \\[1mm]
\rule{0mm}{4mm}
{}~$\varepsilon'_2 \rightarrow \varepsilon_1^{} \varepsilon'_1/\varepsilon_2^{}
$~ & ~$\varepsilon'_2 \rightarrow \varepsilon_1^2/\varepsilon'_2$~ &
{}~$\varepsilon'_2 \rightarrow \varepsilon_2^{}\varepsilon_1^{}/\varepsilon'_1$
\\
\end{tabular}
\end{center}
where $\varepsilon_2=\varepsilon'_1+\varepsilon'_2-\varepsilon_1$.
Under this transformations the sum of integrals over regions I - IV can be
represented as an integral over the region I only:
\begin{equation}
{\int\int}_{(\text{I})} (\varepsilon'_1 \varepsilon'_2 \varepsilon_2)^\alpha
\left[1+\left(\frac{\varepsilon_2}{\varepsilon_1}\right)^\alpha
- \left(\frac{\varepsilon'_1}{\varepsilon_1}\right)^\alpha -
\left(\frac{\varepsilon'_2}
{\varepsilon_1}\right)^\alpha \right]
\left[ 1+\left(\frac{\varepsilon_2}
{\varepsilon_1}\right)^{\beta}
- \left(\frac{\varepsilon'_1}{\varepsilon_1}\right)^{\beta} -
\left(\frac{\varepsilon'_2} {\varepsilon_1}\right)^{\beta} \right]
d\varepsilon'_1 d\varepsilon'_2 \, ,
\label{eq76}
\end{equation}
where $\varepsilon_2=\varepsilon'_1+\varepsilon'_2-\varepsilon_1$ and
$\beta=-7/2+3\alpha$. We see that in fact there are four solutions to the
equation $I_{\rm coll} = 0$ which correspond to $\alpha =0, \,1$ or
$\beta =0, \,1$. This gives
\begin{equation}
\alpha = 0, ~~\alpha = 1, ~~\alpha = 7/6, ~~\alpha = 3/2 ~~.
\label{alfa}
\end{equation}
Our numerical self-similar solution, Fig. \ref{fig6},
tend to the power law $f= \varepsilon^{-7/6}$; however, it never reaches
this critical exponent prior to the condensate formation,  see Figs. 7 and 8.
Physically the solution with $\alpha = 7/6$
corresponds to the constant flux of the particle number from the region of
lager energies towards the future condensate.

\subsection{Solutions in the presence of condensate}
\label{ssec:Icon}

In the presence of the condensate  the  number and the nature of zeros of the
collision integral changes. The kinetic equation can be represented as a
system of two equations, Eqs.(\ref{kin3}). The collision integral in
Eq.(\ref{fdot4}) consists of two terms. The first term, $I_{\rm p} $, is the
same
as in the absence of the condensate, and the corresponding integral is zero on
the solutions corresponding to the same values of $\alpha$,  Eq.(\ref{alfa}).
However, the second term does not. Assuming the
power law $f(\varepsilon) = \varepsilon^{-\alpha}$, the integration in this
term can be made analytically with the result proportional to
\begin{equation}
\varepsilon^{1-2\alpha}
\left[ \frac{4}{\alpha-1}+\frac{4^{\alpha}  \Gamma(1-\alpha)
(\pi + \Gamma(\frac{3}{2}-\alpha)\Gamma(\alpha-\frac{1}{2}))}{2\sqrt{\pi}
\Gamma(\frac{3}{2}-\alpha)}\right].
\label{t23}
\end{equation}
This expression as a function of $\alpha$ has two zeros only at
\begin{equation}
\alpha = 1~~~~\text{and}~~~~  \alpha = 3/2 ~~~~,
\label{alfa1}
\end{equation}
and consequently the total collision integral in Eq.(\ref{fdot4})
is zero on those exponents only. Moreover, the collision integral in
Eq.(\ref{equationa4}) is zero with $\alpha = 1$ only. This explains why the
power law $f(\varepsilon) = \varepsilon^{-7/6}$ changes to $f(\varepsilon) =
\varepsilon^{-1}$ in the presence of the condensate, see Fig. \ref{fig8}b.

\section{Kinetic equation for ideal bosons interacting with degenerate
Fermi gas}
\label{zT}

Assuming fermions to be at zero temperature, {\it i.e.}, their distribution
function to be given by $\eta (\varepsilon) =\theta (\mu -\varepsilon )$,
allows us to make one more integration in
Eq. (\ref{kin2}) analytically. After some algebra we obtain:

\begin{equation}
\frac{df (\varepsilon)}{d\tau} = f(\varepsilon)(J_1 + J_2)+ \lbrack
f(\varepsilon) + 1 \rbrack (J_3 + J_4) \, ,
\label{eq1}
\end{equation}
where
\widetext
\begin{mathletters}
\FL
\begin{equation}
J_1 = -\frac{\eta(\varepsilon)}{\sqrt{\varepsilon}}
\int_0^\varepsilon dx
  x^{1/2}(\varepsilon-x)  \lbrack f(x) + 1 \rbrack ~,
\end{equation}

\FL
\begin{eqnarray}
J_2 &=& \frac{2(1-\eta(\varepsilon))}{3\sqrt{\varepsilon}}  \left\{
\int^\mu_0 dx x^{3/2} \lbrack f(x +\varepsilon-\mu) + 1 \rbrack -
\mu^{3/2}\int_{\mu}^\varepsilon dx  \lbrack
f(x) + 1 \rbrack \right. \nonumber \\
& & \left. - \int_0^{\mu} dx x^{1/2}(\mu-\frac{x}{3})
 \lbrack f(x) + 1 \rbrack \right\}
\label{re1}
\end{eqnarray}

\FL
\begin{equation}
J_3 =  \eta(\varepsilon)\left\{  \int^{\mu}_\varepsilon dx
 (x-\varepsilon) f(x) +
\int^{\infty}_{\mu} dx (\mu - \frac{\varepsilon}{3})f(x) -
\frac{2}{3\sqrt{\varepsilon}}
\int_0^\varepsilon dx
      x^{3/2} f(\mu+\varepsilon-x) \right\}
\label{re2}
\end{equation}

\FL
\begin{equation}
 J_4  = \frac{2(1-\eta(\varepsilon))}{3\sqrt{\varepsilon}}
\left\{    \int^{\infty}_{\varepsilon} dx \mu^{3/2} f(x)
- \int_0^\mu dx
     x^{3/2} f(\mu+\varepsilon-x) \right\}\nonumber\\
\end{equation}
\end{mathletters}
\narrowtext
\noindent
where $\mu$ is the Fermi energy. This equation was subject to
the numerical integration in Section \ref{sec:zeroT}.

\renewcommand{\baselinestretch}{0.5}


\end{document}